\documentclass[3p,final]{elsarticle}
\usepackage{lineno,hyperref}
\usepackage{graphics}
\usepackage{tabularx}
\usepackage{makecell}
\usepackage[caption=false]{subfig}
\modulolinenumbers[5]

\usepackage{soul}

\journal{Journal of \LaTeX\ Templates}

\begin{document}

\begin{frontmatter}
\title{Using Complex Network Theory for Temporal Locality in Network Traffic Flows}

\author[1]{Jin-Fa Wang\corref{mycorrespondingauthor}}
\cortext[mycorrespondingauthor]{Corresponding author.}
\ead{jinfa.wang@mervin.me}

\author[1]{Hai Zhao}
\author[1]{Shuai-Zong Si}
\author[1]{Hao Yu}
\author[1]{Shuai Chao}
\author[2]{Xuan He}

\address[1]{School of Computer Science and Engineering, Northeastern University, Shenyang, China, 110819.}
\address[2]{Sino-Dutch Biomedical and Information Engineering School, Northeastern University, Shenyang, China, 110819.}

\begin{abstract}
Monitoring the interaction behaviors of network traffic flows and detecting unwanted Internet applications and anomalous flows have become a challenging problem, since many applications obfuscate their network traffic flow using unregistered port numbers or payload encryption. In this paper, the temporal locality complex network model--TLCN is proposed as a way to monitor, analyze and visualize network traffic flows. TLCNs model the interaction behaviors of large-scale network traffic flows, where the nodes and the edges can be defined to represent different flow levels and flow interactions separately. Then, the statistical characteristics and dynamic behaviors of the TLCNs are studied to represent TLCN's structure representing ability to the flow interactions. According to the analysis of TLCN statistical characteristics with different Internet applications, we found that the weak interaction flows prefer to form the small-world TLCN and the strong interaction flows prefer to the scale-free TLCN. In the studies of anomaly behaviors of TLCNs, the network structure of attacked TLCNs can have a remarkable feature for three attack patterns, and the evolution of TLCNs exhibits a good consistency between TLCN structure and attack process. With the introduction of TLCNs, we are able to harness a wealth of tools and graph modeling techniques from a diverse set of disciplines.
\end{abstract}

\begin{keyword}
Complex network; Network traffic flow; Temporal locality; Statistical characteristics; Dynamic behaviors
\end{keyword}

\end{frontmatter}


\section{Introduction}

Temporal locality refers to the property that referencing behavior in the recent-past is a good predictor of the referencing behavior to be seen in the near-future\cite{BUDIYONO201440,MAHANTI2000187}. In network traffic flows, the referencing behavior is the interactivity among the flows on network protocol or flow content, that is one current flow may induce the further flows. What method is used to describe the interactive processes of large-scale network traffic flows is the fundamental problem that motivates this paper in the process of monitoring networks. Furthermore, we want to develop a network traffic flow profiling model which is able to determine the application type and detect the anomalous traffic patterns.

The studies of network traffic profiling can be classified by their level of observation: (a) packet level, such as signature-based application detection and methods using the well known port numbers, (b) flow level statistical techniques (c) host level, such as host-profiling approaches. In both the case of malcode and P2P, the above methods using content signature seem destined to fail in the face of encryption and polymorphism, and cannot also think of the flow interactive behavior while it has become the cornerstone of the web caching technique\cite{MAHANTI2000187,JIN2001174}. Therefore, how to present the interaction behavior of network traffic flows is a significant challenge for monitoring the dynamics of Internet traffic.

In this paper, we introduce complex network theory to present the complex interaction behavior of large-scale network traffic flows. Since complex network provides a powerful mechanism for capturing the interactive relationships among study objects, it has been an effective method for relational expression of structured datasets\cite{TANG20134192,WANG2016456,wang201637}, especially the time series data. For instance, in the study of earthquake time series the authors\cite{ABE2005588,HE2014175} developed the earthquake complex network model based on time influence domain, i.e. there is a relation, the former influencing the latter by releasing the energy, between two earthquakes occurring in the same time period. For the road traffic, Zheng et al.\cite{ZHENG20086177} proposed a simple weighted network model which presents the similarities of traffic flow states by defining the traffic flow state as network node. However Tang et al. used complex network theory to study the similarities of sampling points on time series \cite{YAN2017149,TANG20132602,TANG2014303}, and to study the complexity of network states extracted from traffic flow time series\cite{TANG20134192,TANG2016635}. Zanin\cite{ZANIN2014201} constructed the complex network representing of air traffic flows to identify the situations in which probability of appearance of Loss of Separation events is increased. For Internet, based on complex networks theory the complexity of Internet topologies have been widely studied \cite{Faloutsos1999,FAN2016327,CHEN2018191}, but there is very little research about the complexity of traffic flows. In 2007, a study of the social behavior of Internet hosts is presented by Marios et.al \cite{Iliofotou2007}, who firstly proposed the Traffic Dispersion Graphs (TDGs) as a way to monitor, analyze, and visualize network traffic. In TDG, the edge can be defined to show different interactions between two hosts. Subsequently, Wang et.al \cite{Xu2014} studied social behavior similarity of Internet end-hosts based on behavioral graph analysis. The above two works have archived outstanding results in term of representing the interactions or social behaviors among the hosts. In this paper, a novel model--temporal locality complex network (TLCN) is proposed by focusing on the interaction behaviors of network traffic flows on temporal locality, e.g. the dependency or correlation among the flows. First, we construct complex networks of traffic flows based on temporal locality, and give node filtering strategies and edge filtering strategies aimed at different packet levels. Then, the relationships between temporal locality window and traffic throughput, between temporal locality window and TLCN structure are analyzed by constructing different temporal-scale TLCNs. Finally, the TLCNs are used to study the complex behaviors of large-scale network traffic flows. On one hand, the analyses of statistical characteristics of static TLCNs constructed based on Internet application help us to discover the TLCNs' structure representing ability and application behaviors associated with the protocols. On the other hand, the studies of the structure evolution of the TLCNs with anomaly events (e.g., short-term attack and long-term botnet) are to explore the dynamic behaviors, find the attack patterns and detect the anomalous traffic.

The rest of this paper is organized as follows. In Section 2, the temporal locality complex network is proposed to describe the interaction behaviors of traffic flows. Section 3 is devoted to study the statistical characteristics of static TLCNs based on Internet application. In section 4, the dynamic behaviors of the TLCNs with anomaly events are analyzed. And we discuss our findings in Section 5. Finally, Section 6 presents the conclusions of this paper.

\section{Temporal locality complex network}
\label{sec:temporal-locality-complex-network}
Temporal locality is used to describe the dependency or correlation on a per item basis apart from in an aggregate reference flow\cite{MAHANTI2000187}. Here a specific example of flow interactivity, once Google search, is given, as a fragment of flow traces is shown in Fig.\ref{sfig:net-trace}. When we are going to open the Google.com, the browser firstly looks up the IP address of Google.com by the DNS protocol flows, and loads Google's webpage by HTTP protocol flows. Then, it sends the HTTP request flows again to obtain the search results for the given keywords. Next, we probably click some hyperlinks of the search page to trigger new flows. Eventually, we may repeatedly request new flows decided by the content of previous flow, until we obtain our wanted results. Obviously, the flows in each step depend on the previous flow, that is to say the previous flows induce the further flows. Another example is the recursive or iterative query of DNS resolver. The DNS server has to forward this requests to its provider or tell the user its provider, if there is no record for a given domain name\cite{Mockapetris1988}. Observing the flow traces, it can be seen that in the vast majority of cases one flow triggers a few of further flows. In other words, the dependency relationships are built from current flows to further flows. In fact, the interactions in the large-scale traffic flows are so complicated that the traditional statistical methods are very hard to represent the macroscopic relationships. So in this paper the complex network theory is introduced to describe the complexity and interactivity of large-scale traffic flows. Meanwhile, using complex network to temporal locality of network traffic flows is a novel method in term of describing the interaction of network traffic flows and exploring the flow behaviors.

\subsection{Definition}
\label{sec:definition}

First, a 5-tuple $f = \{si, di, sp, dp, p \}$ is defined to represent one network traffic flow, where the notations $si$, $di$, $sp$, $dp$ and $p$ denote the source IP, destination IP, source port, destination port and protocol number, respectively. Then the set $F = \{ f_{1}, f_{2}, \dots, f_{n} \}$ denotes network traces of $n$ traffic flows on time series, as shown in Fig.\ref{sfig:net-trace}. If the $si$, $di$, $sp$, $dp$ and $p$ of two flows are same separately, they should belong to one flow. When we focus on the traffic flows of one Internet application (i.e., either of two ports is a fixed value), the 5-tuple can be simplified into the 2-tuple $f=\{si, di\}$, as shown in Fig.\ref{sfig:net-construction}. Here one unique flow can be identified by the values of $si$ and $di$. For two flows $f_i$ and $f_j$ occurring at time $t_i$ and $t_j$ separately ($t_i \leq t_j$ ), if $t_j \in [t_i, t_i + \Delta w]$, there is the interactive relationship from $f_i$ to $f_j$, where the $\Delta w$ denotes the length of temporal locality window. In other words, there is a directed connection from flow node $f_i$ to $f_j$, or vice versa. Furthermore, the network traffic flow time series $F$ are translated into one complex network $G=(V, E)$, where the $v \in V$ is the network node that denotes unique network flow $f$, the $e \in E$ describes a dependency relationship between two nodes. In this paper the $G$ is a directed complex network, where the direction of edge $e(v_i, v_j)$ indicates the flow $v_i$ triggers the flow $v_j$ or the $v_j$ depends on the $v_i$.

Based on above definition, the temporal locality complex network (TLCN) can be constructed from the network traffic traces. About TLCN, the most important is the ability of describing macroscopic dependent or correlative structure, even though it builds the connections from microscopic traffic flows. Therefore, in the IP network one TLCN node $v$ denotes a unique flow entity $f$ with distinct source IP, destination IP, source port, destination port or protocol, and the TLCN can capture the interactivity and dynamic of network traffic flows.

\subsection{TLCN Filtering}
\label{sec:network-filtering}

One of the fundamental questions in using TLCN is the definitions of network node and edge. This basic question can be answered in many different ways depending on the goal of our study. We start with the observation that what kind or level of network flows should be selected as network node in TLCN. We call this process \textit{Node Filtering}. One simple node filtering is to select the IP protocol flows. In addition to this basic node filtering, we can enrich the definition of what constitutes a node by imposing "stricter" rules that capture different aspects of traffic flows. For instance, we can have filters for "allowing" a flow node based on: (a) the frequency of one flow, (b) the type of the flow (e.g., TCP three-way handshake), (c) the application protocol used (TCP, UDP, ICMP etc.), (d) the application based on port number (e.g., Port Number 80 for HTTP, Port Numbers 6881$\sim$6889 for BT), and finally (e) looking at properties of the flow content, such as payload size or by using deep packet inspection.

Besides basic definition about network edge in section \ref{sec:definition}, more rules or features, called \textit{Edge Filtering}, can be put forward to enrich the definition of network edge. In general, the directed edges can be used to identify the indicator of the probability interaction between a pair of flows. As we will later see, directed edges in a TLCN are very useful in identifying various node behaviors and also in establishing their causal relationship. However, we could choose to consider undirected edges, which will enable us to use the more extensively studied complex network metrics for undirected networks, as discussed in later sections. In addition to edge direction, it is also important to define the level of network edge in TLCN. One simple edge filter is to add an edge $e(v_i,v_j)$ between flow node $v_i$ and $v_j$ when the flow $v_j$ is observed in the temporal locality window $\triangle w$ of the flow $v_i$. Once an edge is added, this filter ignores any flow interactivity from $v_i$ to $v_j$. We call this edge filter as the \textbf{Unweighted-Edge(UWE)}, and is mainly used to study the interactive process of network flows. However, for the flow interaction behavior, the frequency of edge $e(v_i,v_j)$ is an important indicator in TLCN. We call this edge filter as the \textbf{Weighted-Edge(WE)}.

\subsection{TLCN Formation}
\label{sec:tlcn-formation}

In this paper, we mainly focus on the interactions of the traffic flows based on Internet applications including ICMP application protocol-based TLCN using the (c) filtering type and application port-based TLCN using the (d) filtering type (as defined above). Throughout the paper and unless stated otherwise, when the legacy application for a port uses TCP or UDP, we use the \textbf{WE edge filter} on the corresponding source or destination port of the flows (e.g., TCP Port 25 for SMTP, UDP Port 53 for DNS). For ease of presentation, we will refer to each TLCN using the name of the dominant or well-know application under that port. For instance, the HTTP TLCNs is formed by using all the TCP and HTTP flows that have as source or destination port the number 80.

\begin{figure}[htbp]
\begin{center}
\subfloat[A fragment of network traffic traces.\label{sfig:net-trace}]{
    \includegraphics[width=0.4\columnwidth]{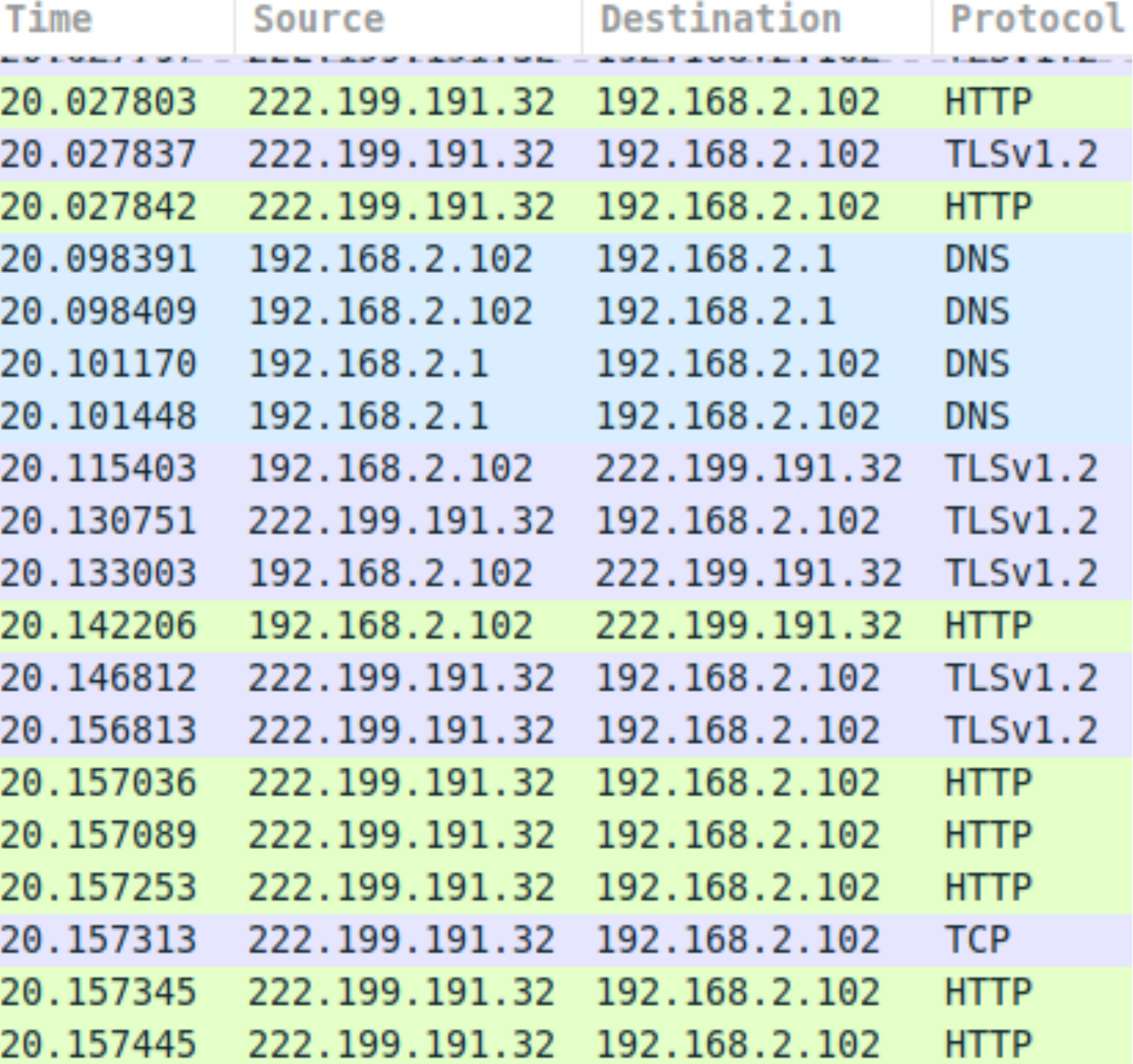}
}\hfill
\subfloat[Diagram of TLCN construction of a network application flows.\label{sfig:net-construction}]{
    \includegraphics[width=0.55\columnwidth]{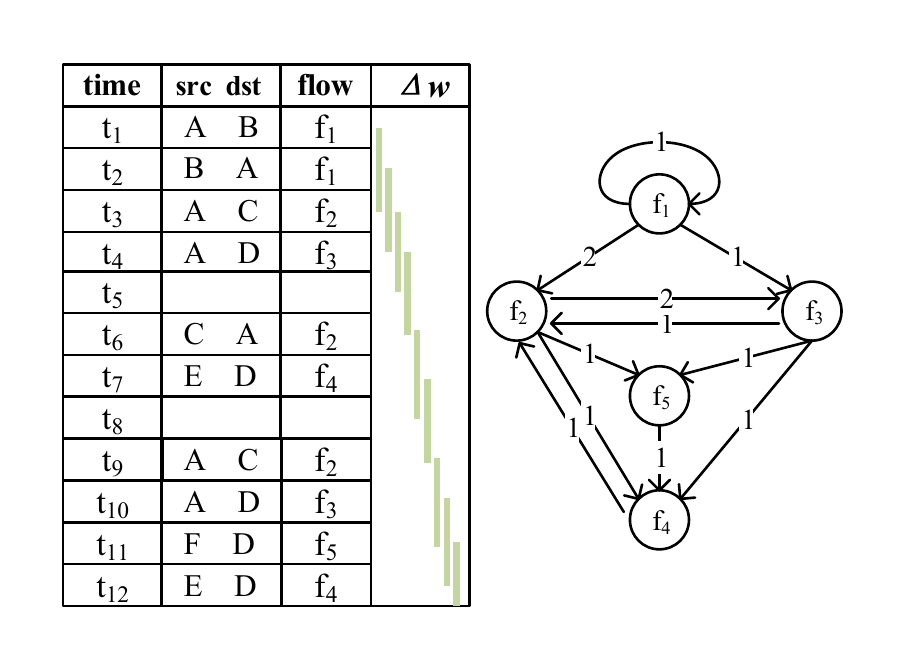}
}
\caption{TLCN construction from network traffic traces. (a) The real network traces for a once search, (b) the right table presents the 10 2-tuple traffic traces captured from 12 time units where the temporal locality window $\Delta w$ is 3 time unit. The left is a constructed directed weighted TLCN corresponding to the right traffic traces.}
\label{fig:tlcn}
\end{center}
\end{figure}

Since we use node filtering by port number and edge filtering by the edge frequency, as shown in Fig.\ref{sfig:net-construction}, the TLCNs capture aspects of any application that uses these ports. However, application port-based filtering is consistent with our use of TLCNs as a monitoring tool. For example, if at some time points network traffic at TCP Port 80 appears significantly different, it could be: (a) a new begin or malicious application tunneling its traffic under that port, or (b) a change in the behavior of the traditional traffic.

\subsection{Dataset and temporal locality window} 
\label{sec:dataset-tlw}
In order to study the structure characteristics and analyze the dynamical behaviors of TLCNs, we use a variety of publicly available real Internet traffic traces, such as WIDE backbone traffic\footnote{\url{http://mawi.wide.ad.jp/mawi/}}, CTU botnet traffic\footnote{\url{https://www.stratosphereips.org/datasets-ctu13/}}, and DARPA traffic\footnote{\url{https://www.ll.mit.edu/ideval/index.html}}. These traces are non-sampled and include up to layer-4 headers with no payload. The WIDE data is collected by tracing at the transit link of WIDE to the upstream ISP. So the WIDE dataset represents a normal backbone network flows\cite{Fontugne2010}. The CTU botnet traffic dataset was captured in the CTU university in 2011, whose goal was to have a large capture of real botnet traffic by executing the Neris malware which mixed with normal traffic and background traffic\cite{Golkar2014}. The Darpa99 is a richer dataset that mixes with five major attacks (i.e, DoS, Probe, R2L, U2R and Data attack) and background traffic on the tested network. As consequence, Table \ref{tab:dataset} gives some details on three datasets including the start time, duration of captured traffic, the number of unique IPs, the number of 2-tuple flows and the number of 5-tuple flows.

\begin{table}[htbp]
\begin{center}
\caption{Our set of publicly available network traffic from WIDE, CTU and Darpa. The 2-tuple flow only includes source and destination IP, but the 5-tuple flow includes source and destination ports, and protocol except for source and destination IP.}\label{tab:dataset}
\scalebox{0.9}{
\begin{tabular}{ c  c  c  c  c  c }
\hline
\hline
Name & Time  & Duration & Unique IPs & 2-tuple flows & 5-tuple flows\\
\hline
WIDE        & 2018-01-10 14:00:00 & 15 ms & 14,891,611 & 18,003,031& 23,630,244  \\
CTU-50      & 2011-08-17 12:01:01 & 5.18 hs & 367,215 & 457,131 & 2,300,253  \\
Darpa   & 1999-04-01 08:00:01 & 110 hs & 2,413 & 7184 & 410,147  \\
\hline
\hline
\end{tabular}
}
\end{center}
\end{table}

In the TLCN, temporal locality window $\Delta w$ is the only parameter to determine whether there is a connection between two flow nodes or not. Thus how to select an appropriate temporal locality window is very important for TLCN structure. Theoretically, the bigger the $\Delta w$ is, the more out-connections one flow node would have. The network density of the TLCN is higher accordingly. For this reason, the $\Delta w$ is large enough that the TLCN loses the purpose of depicting the dependency or correlation relationship among the flows. In fact, the temporal locality of network traffic flows is the referencing behavior in short time. But, if it is too small, some valuable connections in TLCN would be filtered. Thus the $\Delta w$ should be determined to a proper value by which the complex network TLCN can capture the characteristic of the network traffic flow time series. In the below, the temporal locality window selection is studied from two aspects: network structure and network throughput.

\begin{figure}[htbp]
\captionsetup[subfigure]{labelformat=empty}
\subfloat{%
  \includegraphics[width=.49\columnwidth]{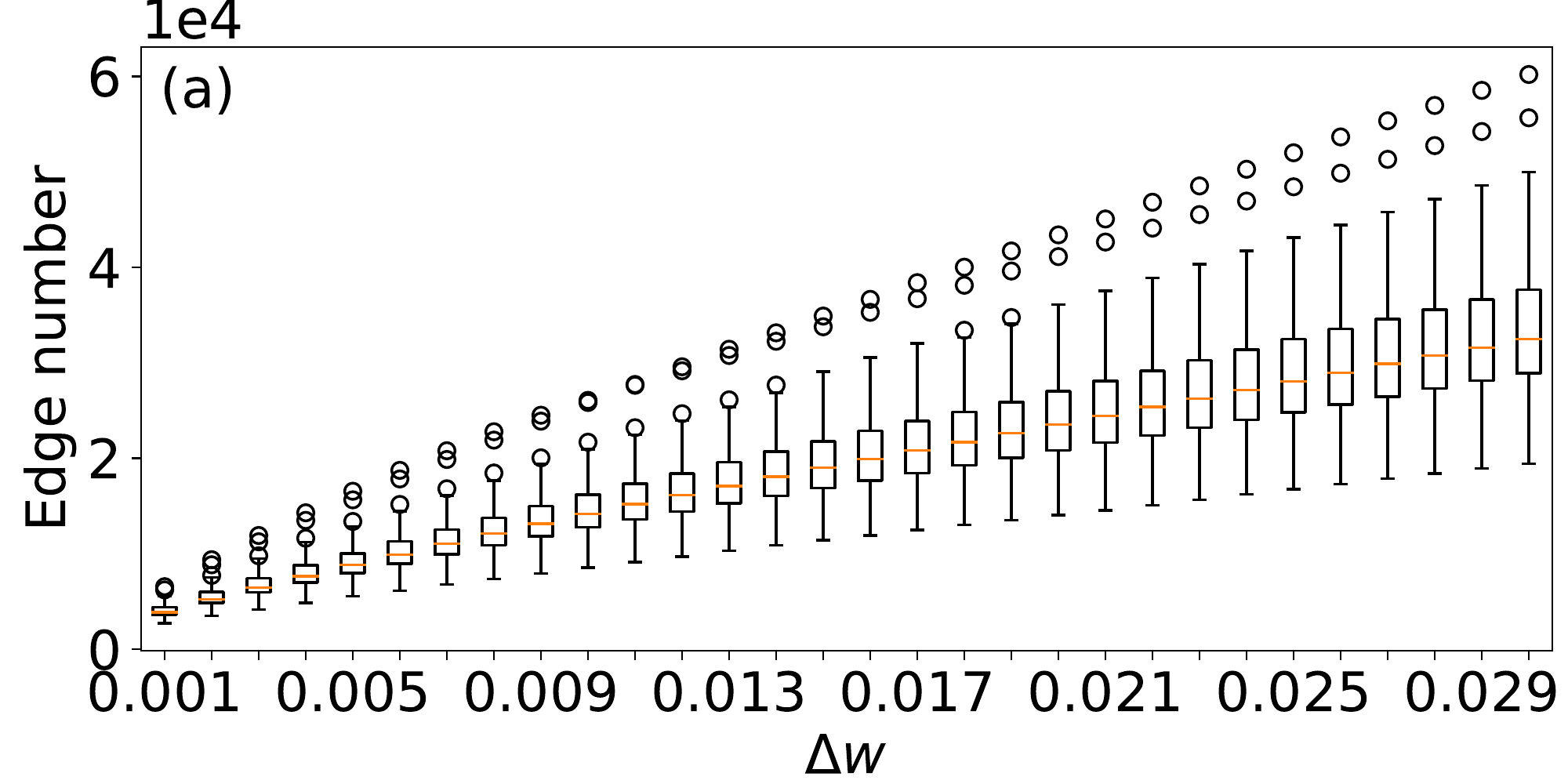}%
  \label{sfig:wide-53-w-edge}
}\hfill
\subfloat{%
  \includegraphics[width=.49\columnwidth]{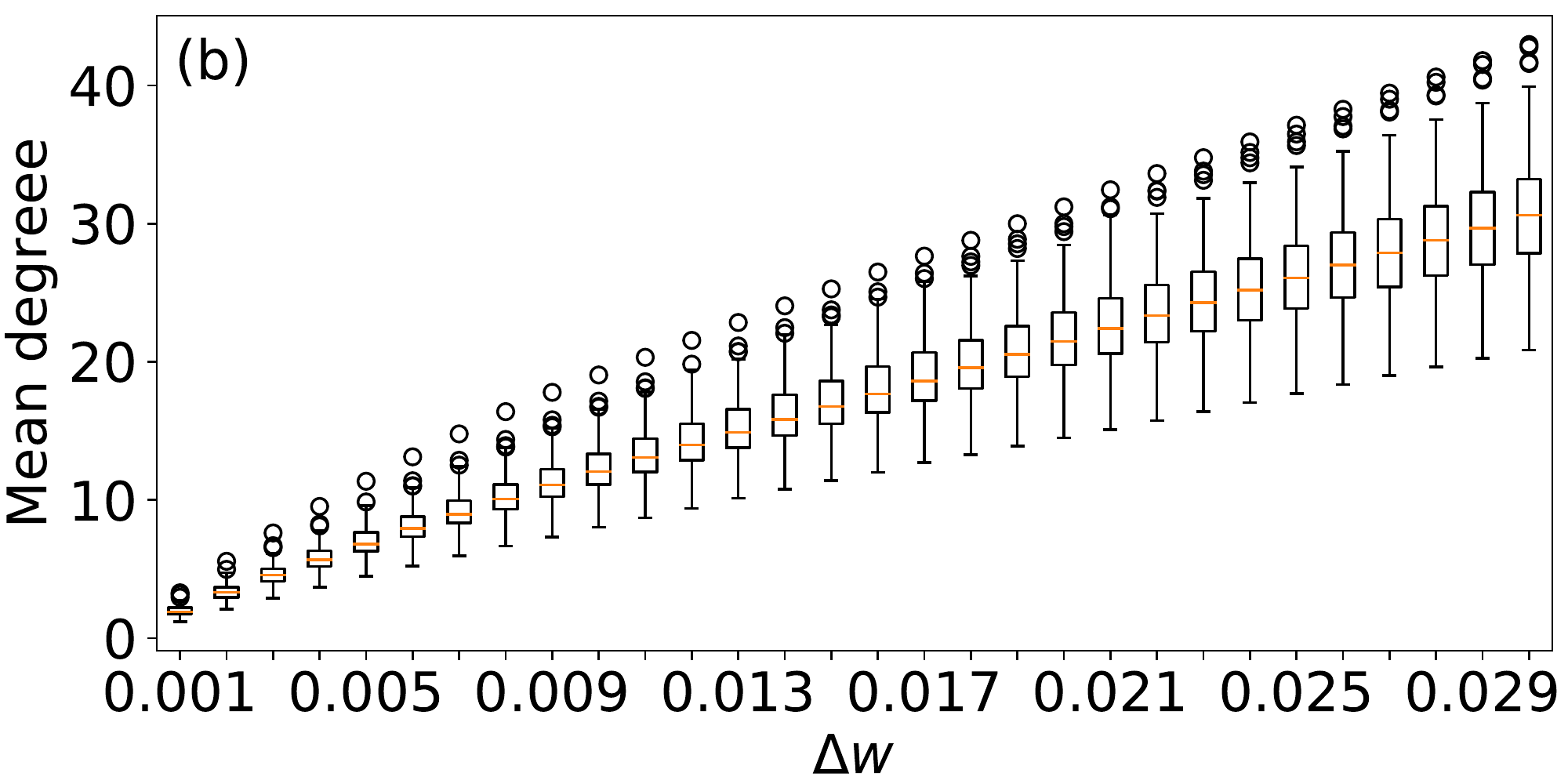}%
  \label{sfig:wide-53-w-md}
} \\
\subfloat{%
  \includegraphics[width=.49\columnwidth]{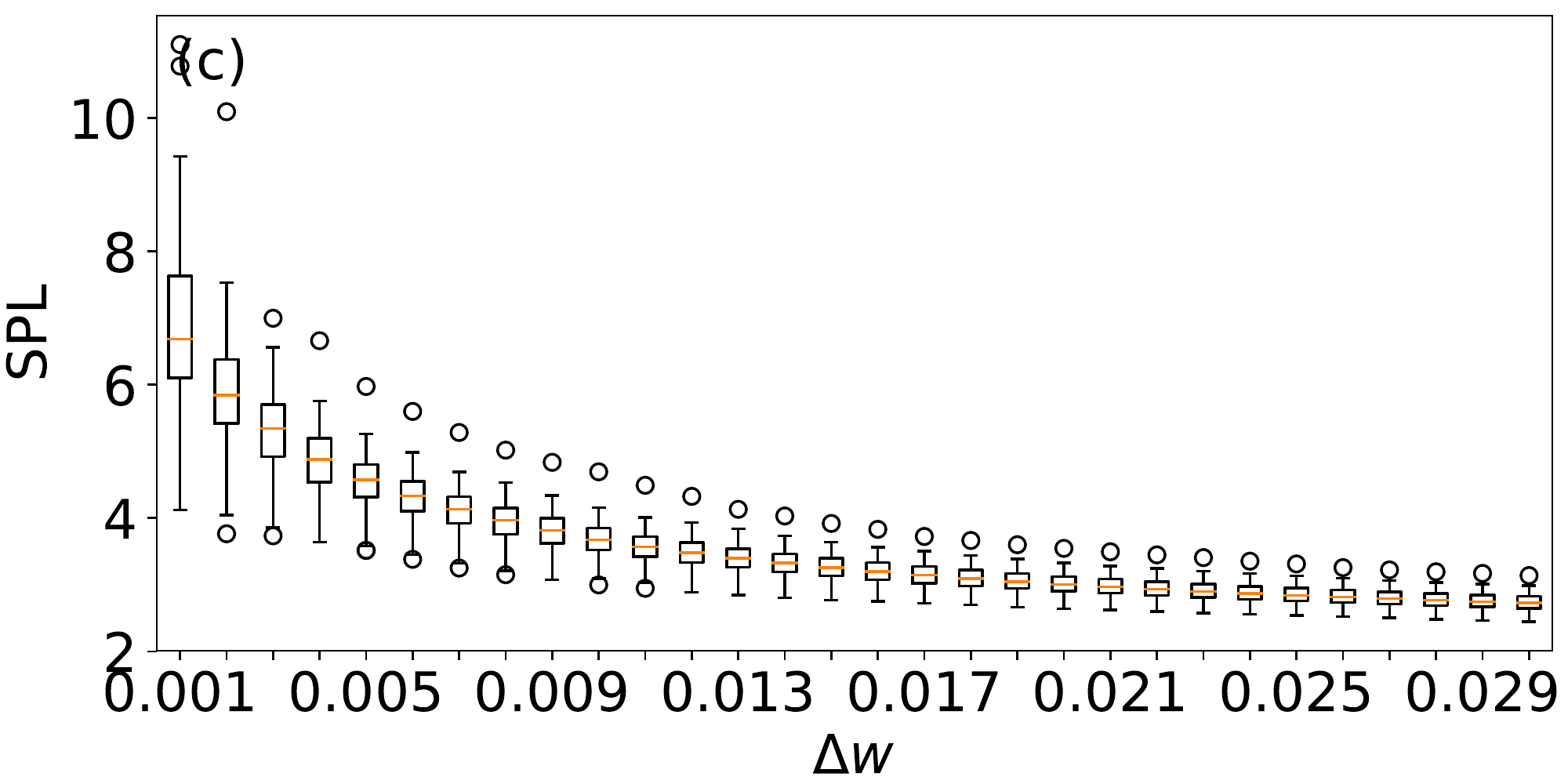}%
  \label{sfig:wide-53-w-spl}
}\hfill
\subfloat{%
  \includegraphics[width=.49\columnwidth]{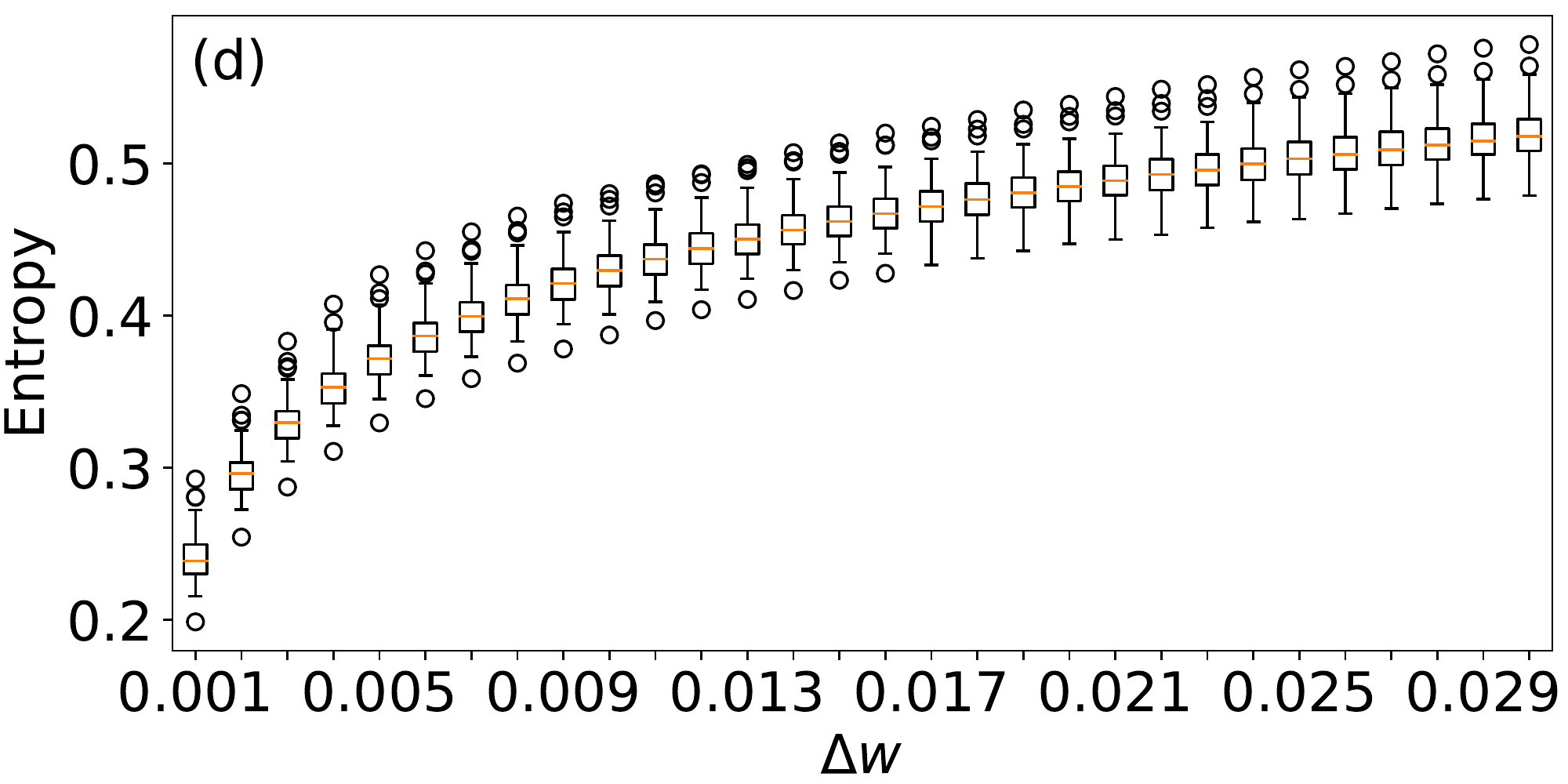}%
  \label{sfig:wide-53-w-entropy}
}
\caption{Distribution between the characteristics values of DNS-based TLCNs extracted from WIDE dataset and temporal locality window $\Delta w$. (a) edge number vs. $\Delta w$, (b) mean degree vs. $\Delta w$, (c) shortest path length (SPL) vs. $\Delta w$, and (d) entropy vs. $\Delta w$.}
\label{fig:w-all}
\end{figure}

On the one hand, we study the correlation between the value of $\Delta w$ and TLCN structure by constructing the DNS application-based TLCNs with different $\Delta w$. To begin with, the 15 minutes of WIDE traffic traces is divided into 90 groups with a 10-second sampling interval. Given a temporal locality window $\Delta w$, each group of traffic traces can be translated into one TLCN. Then using the boxplot, as shown in Fig.\ref{fig:w-all}, is to plot the mean, minimum, and maximum of a characteristic sequence which represents the characteristic values of 90 TLCNs with same $\Delta w$. According to the distribution of the number of edge, mean degree, SPL and entropy as a function of $\Delta w$ in Fig.\ref{fig:w-all}, two linear correlation are found between the network edge number and the temporal locality window $\Delta w$, between mean degree and $\Delta w$. Just like discussed above that the higher $\Delta w$ makes one flow node cover more future flow nodes, i.e, build the connections with more nodes, but the number of network nodes is a known value with a fixed sampling interval. Therefore there is a linear correlation between one-order TLCN metrics and the $\Delta w$. Additionally, the more network edges improve the density of TLCN and shorten the shortest distance of some pairs of nodes\cite{BAGLER20082972}. The network shortest path length (i.e., $SPL=\frac{1}{n(n-1)}{\sum_{i \neq j}{d(v_i,v_j)}}$, let $d(v_i, v_j)$ denote the shortest distance between node $v_i$ and $v_j$) decreases exponentially with the increasing value of $\Delta w$. In the Fig.\subref*{sfig:wide-53-w-entropy}, it can be found that the entropy of TLCNs increases power-exponentially with the increasing $\Delta w$. If the $\Delta w$ is close to the sampling interval, the network entropy would be close to 1 owing to the growing network arriving the fully-connected disordered state. As we all know, the higher the systemic entropy is, the harder it is to accurately describe its microscopic state\cite{Crooks1999}. So it is also important to analyze the relationship between the network structure and the application flow behaviors that constructs the complex network from the time series with an appropriate entropy.

\begin{figure}[htbp]
\begin{center}
\includegraphics[width=0.6\columnwidth]{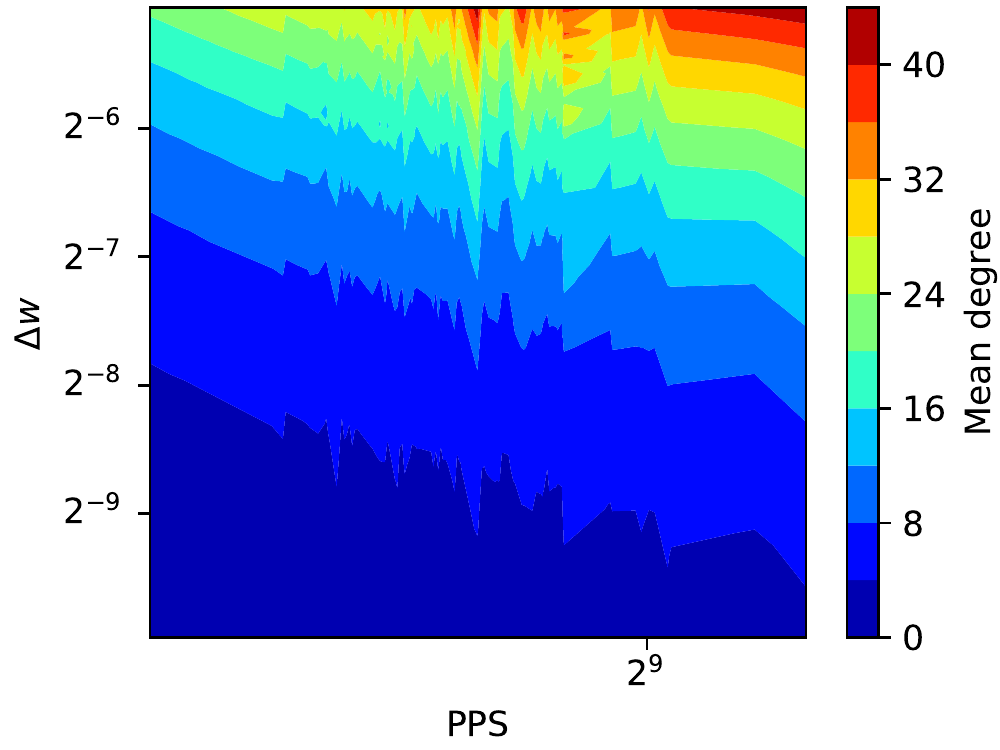}%
\caption{Distribution among network throughput (PPS), temporal locality window ($\Delta w$) and network mean degree ($<k>$) of DNS-based TLCNs extracted from WIDE dataset.}
\label{fig:pps-w-md}
\end{center}
\end{figure}

On the other hand, we study the correlation between temporal locality window $\Delta w$ and network traffic throughput. The Fig.\ref{fig:pps-w-md} shows the distribution among network throughput (PPS), temporal locality window $\Delta w$ and network mean degree($<k>$). In general, the higher the network throughput, the more out-connections one flow node should build. This result is verified by the distribution between PPS and mean degree yet, when we observe it with the fixed $\Delta w$ of the Fig.\ref{fig:pps-w-md}. In other word, the higher throughput improves the TLCN's network density. When the network mean degree is fixed, we found there is a linear correlation between network throughput and $\Delta w$. It is inferred that the network throughput has same feature with the temporal locality window for our TLCN model.

\subsection{TLCN Visualization}

Traditionally, visualization of network traffic in monitoring tools has largely been limited to visualizing measures of traffic volumes on a per flow basis. By contrast, we show that TLCNs lend themselves to simple graphical visualization of flow interaction patterns. For instance, the networks in Fig.\ref{fig:net-all} show a simple set of TLCNs, where we filter the nodes for distinct application Port Numbers (i.e., the Fig.\ref{sfig:net-telnet} $\sim$ Fig.\ref{sfig:net-dns}) and the ICMP application protocol(Fig.\ref{sfig:net-icmp}). Studying these TLCNs (Fig.\ref{fig:net-all}), we can quickly reach the following conclusions, which we have corroborated with a number of similar visualizations:
\begin{enumerate}
    \item \textbf{TLCNs are not a single family of complex networks.} We can see that TLCNs have significant visual and structural differences, which we quantify with network metrics in Table \ref{tab:wide-result}. This characteristic is what gives TLCNs descriptive power, as we discuss later in more detail.
    \item \textbf{TLCNs capture many interesting patterns of network traffic flow interactions.} We can identify several distinctive structures and patterns in TLCNs, which are indicative of the behavior of different applications.
    \begin{enumerate}[a)]
      \item \textit{Node degrees} - The degrees of various nodes and their connectivity in a TLCN help us in visually determining the type of relationship among the nodes. In general, the TLCNs corresponding to applications with a prevalence of flow interactions, such as SMTP (Fig. \ref{sfig:net-smtp}), BT (Fig. \ref{sfig:net-bt}) are dominated by a few high degree nodes whereas the degrees of the TELNET (Fig. \ref{sfig:net-telnet}), DNS (Fig. \ref{sfig:net-dns}) and ICMP (Fig. \ref{sfig:net-icmp}) are fairly well-distributed.
      \item \textit{Node chain} - A directed node chain shows the dependency direction of each of nodes on a chain. For instance the node chain $a \to b \to c$ denotes the flow direction in which the $a$ triggers $b$ and then the $b$ triggers the $c$. Thus the long node chain is more meaningful and visual in aspects of the flow behaviors. Admittedly, the strong-interaction flows prefer to form node chain, such as the HTTP that we step by step to click the hyperlinks on the web pages to obtain our wanted content, but some weak interaction flows are also possible to build node chain due to the ruled scan behaviors (e.g. the TELNET in Fig.\ref{sfig:net-telnet}).
      \item \textit{Nodes community} - There are many community structure whose nodes closely connect to each other. Compared with the node chain presenting the dependency direction of the nodes, the community structure suggests that there are well interdependencies among the nodes in a community, such as the TELNET, BT, ICMP. In the section \ref{sec:statistical-characteristics-of-static-tlcns}, we find that the Internet scan flows are easier to form nodes community.
    \end{enumerate}
\end{enumerate}
\textit{Discussion:} Although our goal here is to visually examine the various properties of TLCNs, good visualization methods have their own value. In fact, effective visualization and human monitoring can often be a more viable alternative to complicated automated methods for anomaly detection. While visualization is useful by itself, if TLCNs are to be used for application monitoring, it is important to translate visual intuition into quantitative statistic.

\section{Statistical characteristics of TLCNs}
\label{sec:statistical-characteristics-of-static-tlcns}

In this section, the statistical characteristics of static TLCNs are studied to find the correlation between TLCN structure and Internet application. First, 6 types of TLCNs constructed based on Internet applications, i.e., TELENT(Port Number: 23), STMP(Port Number: 25), BT(Port Numbers: 69001-69009), HTTP(Port Number: 80), DNS(Port Number: 53), and ICMP, were extracted from the 15 minutes of WIDE traffic traces. As shown in Table \ref{tab:net-params}, the first three rows give the traffic size, traffic fraction and throughput of each of Internet applications. Then, the traffic traces of each of applications are divided into 90 groups with the 10-second sampling rate. Accordingly, the sampling traffic flows of every application are translated into a TLCN with the temporal locality window $\Delta w$ determined by the mean degree. Here the TLCNs are selected with its mean degree being in $[10, 13]$, as shown in Table \ref{tab:net-params}.

\begin{figure}[htb]
\subfloat[TELNET\label{sfig:net-telnet}]{%
  \includegraphics[width=.33\columnwidth]{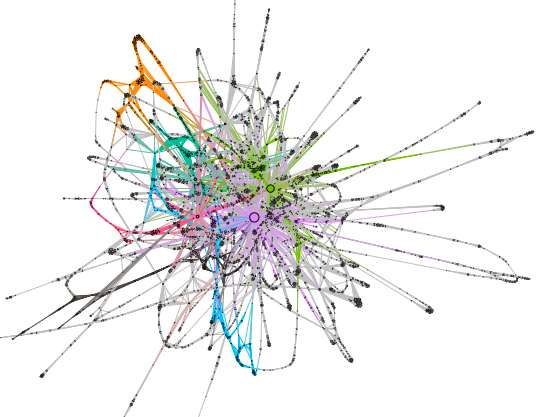}%
}\hfill
\subfloat[SMTP\label{sfig:net-smtp}]{%
  \includegraphics[width=.33\columnwidth]{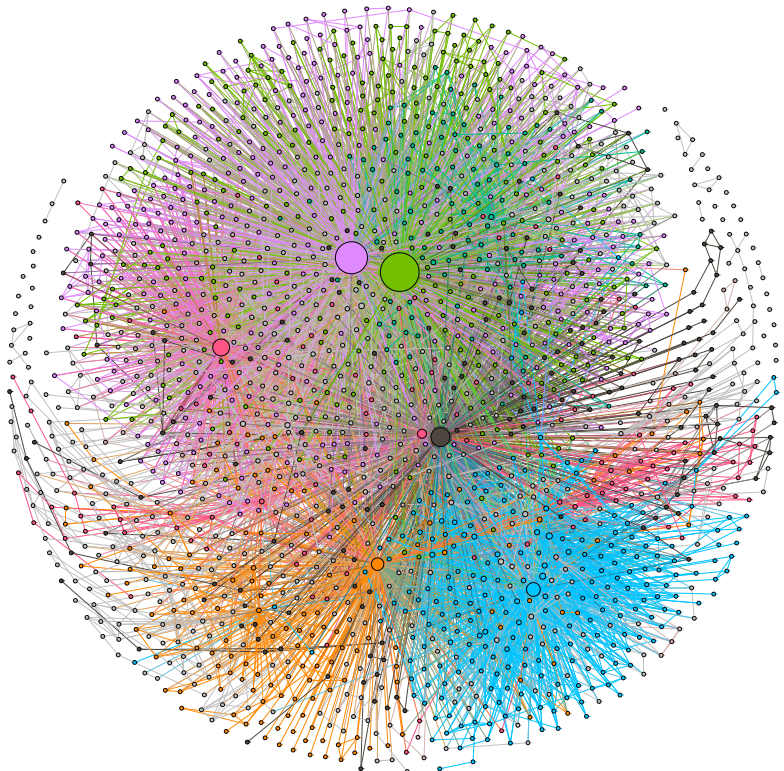}%
}\hfill
\subfloat[BT\label{sfig:net-bt}]{%
  \includegraphics[width=.33\columnwidth]{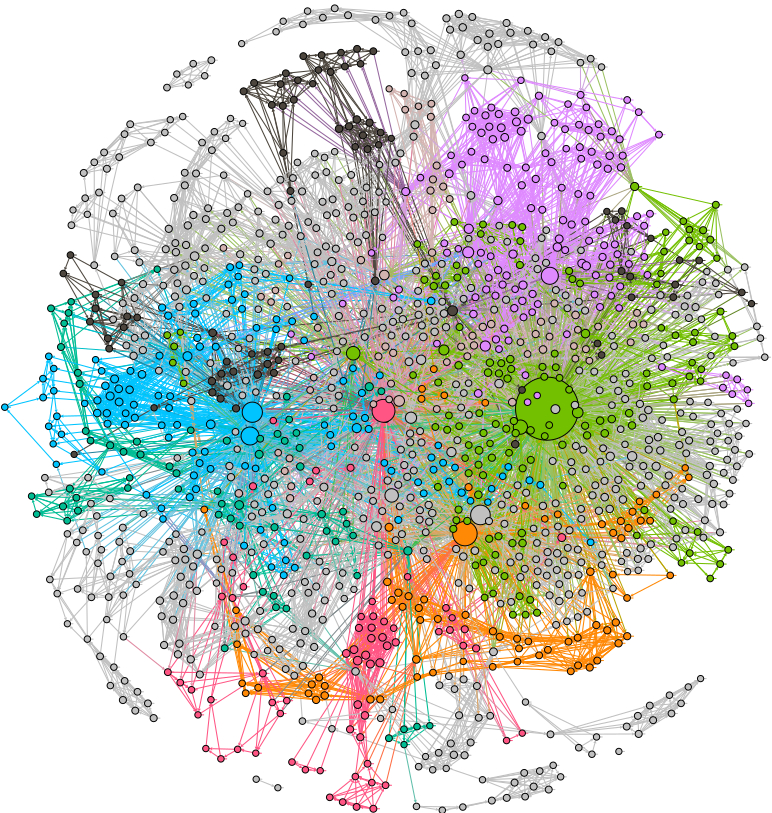}%
}\\
\subfloat[HTTP\label{sfig:net-http}]{%
  \includegraphics[width=.33\columnwidth]{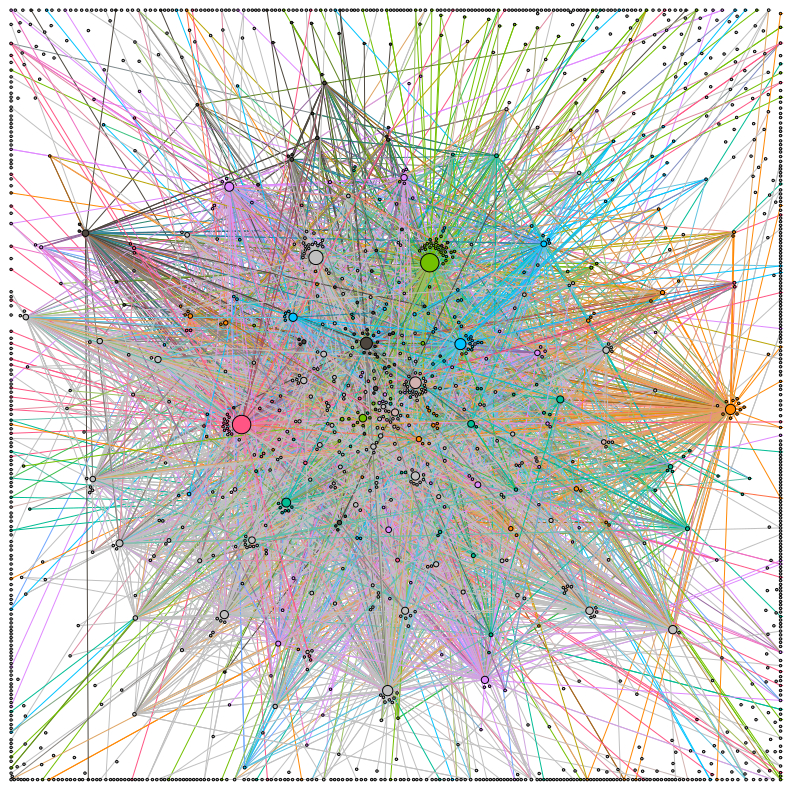}%
}\hfill
\subfloat[DNS\label{sfig:net-dns}]{%
  \includegraphics[width=.33\columnwidth]{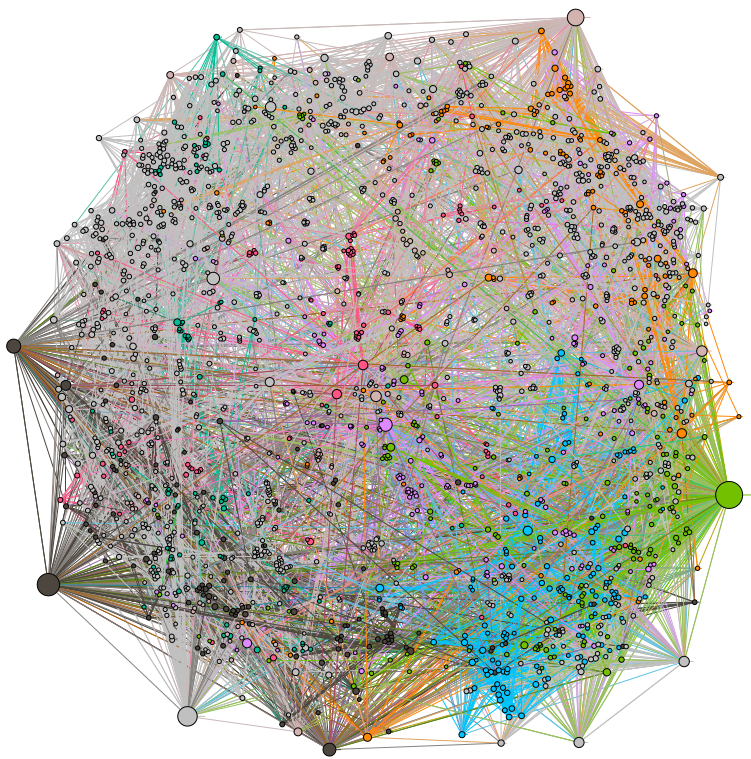}%
} \hfill
\subfloat[ICMP\label{sfig:net-icmp}]{%
  \includegraphics[width=.33\columnwidth]{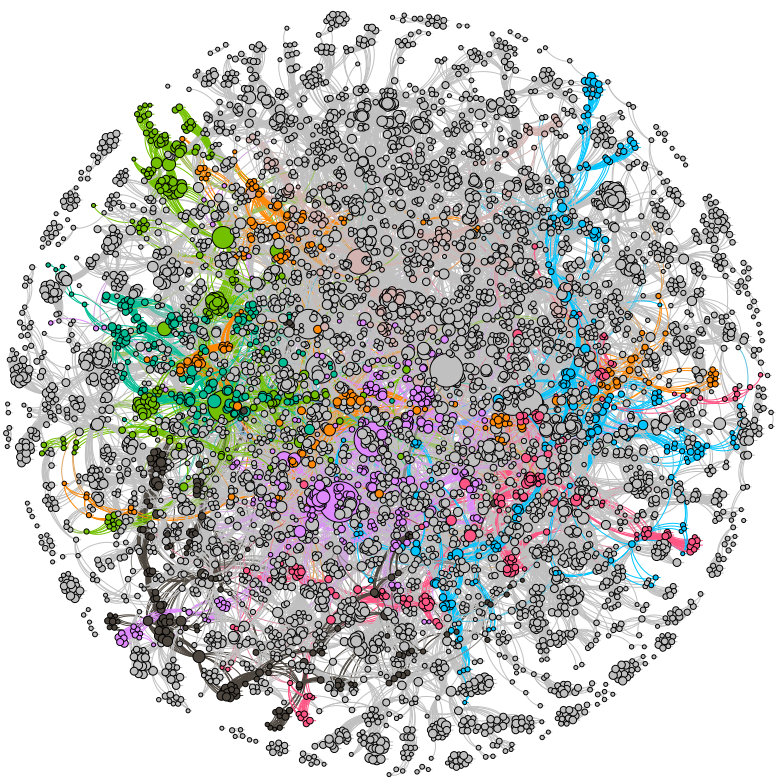}%
}
\caption{TLCN visualization for various applications from WIDE dataset. The size of network node denotes node's out-degree, and the color indicates different community structure in the TLCN.}
\label{fig:net-all}
\end{figure}

Furthermore, Table \ref{tab:wide-result} shows that some network characteristics was calculated. Noted that the values are also averaged over the number of sampling TLCNs, and the values in the parenthesis provide the standard deviations of each network characteristic, which are typically small. The small standard deviations suggest that:(a) the TLCN characteristics seem very stable over the sampling interval, and (b) that temporal locality window $\Delta w$ is a reasonable interval of constructing the TLCN formation. The network characteristic metrics include maximum degree ratio (MDR), clustering coefficient, rich club coefficient, degree distribution and assortativity coefficient, shortest path length and network diameter in which the MDR presents the importance of one central node, the clustering coefficient and rich club coefficient describe the node connectivity, the degree distribution and assortativity coefficient reveal the connection preference, and the shortest path length and network diameter quantify the network performance of information transmission. Additionally, network visualizations of 6 types of TLCNs are given in the Fig.\ref{fig:net-all}.

\begin{table}[htbp]
\begin{center}
\caption{Statistical characteristics of traffic flows and corresponding TLCN of each of Internet applications, and the value of the $\Delta w$. The notations $N$, $r$ and PPS denote the traffic size, traffic fraction and packets per second of a specific application (i.e., network throughput). Given the temporal locality window $\Delta w$, mean degree $<k>$ and entropy $e$ of sampling TLCNs are calculated.}
\label{tab:net-params}
\resizebox{0.9\textwidth}{!}{
\begin{tabular}{ c | c  c  c  c  c  c }
\hline
\hline
network   & TELNET  &  SMTP   &  BT     &  HTTP     &  DNS    &  ICMP \\
\hline
$N$     & 919,614 & 130,804 & 11,390  & 22,568,915 & 381,396 & 19,802,194 \\
$r(\%)$ & 0.77    &  0.11   &  0.01   &  19.67     &   0.33  &  17.26 \\
PPS     &  1022  &   145   &    13  &  25077  &   424    &   22002 \\
\hline
$\Delta w$(s) & 0.003  &  0.005   &  0.01   &  0.00001    &   0.008  &  0.0002  \\
$<k>$   &  11  &  10   &  12   &  11     &   11  &  13 \\
$e$     &  0.3365  &  0.3622  &  0.4052   &  0.3310    &   0.4147  &  0.3836 \\
\hline
\hline
\end{tabular}
}
\end{center}
\end{table}

\textbf{Max Degree Ratio (MDR):}\label{sec:mdr}
MDR is the maximum degree in the network normalized by the number of nodes minus one (i.e., the maximum possible degree of a node in the network). Intuitively, high MDR suggests the presence of a dominate flow node in the TLCN, e.g., the communication between two busy servers. As shown in Table \ref{tab:wide-result}, SMTP usually has at least one very high degree node indicating the presence of high frequency mail transmission flows between two mail servers. Additionally, the BT as one of the most important file share applications also has the high degree nodes. It implies there is a large number of file transmission flows between two hosts. Maybe the visualizations of SMTP (Fig. \ref{sfig:net-smtp}) and BT (Fig.\ref{sfig:net-bt}) are more vivid for the MDR nodes in which the degrees of these nodes are depicted by the cycle size. It can also be seen that the standard deviations of the MDR of SMTP and BT are higher than others. The results suggest that the traffic flows of SMTP and BT are non-continuous and unstable compared with the others. Frankly, this is because of th outbreak or abnormal flows in the monitored network, such as sudden large file transmission on SMTP or BT, or the intentional attack. The abnormal flows, especially the end-to-end attack, will induce or trigger the more other flows in temporal locality window. Just like the Fig. \ref{sfig:net-dos-s-s}, the largest node with red color represents the DoS attack flow from single attacker to single victim.


\textbf{Clustering coefficient:}\label{sec:clustering-coefficient}
Local clustering coefficient $c(i)$ quantifies how well connected are the neighbors of a node $i$ in the network. The network clustering coefficient $C$ denotes this transitivity of all nodes. In the TLCN, the high $c(i)$ indicates there is a well connection among the neighbors of node $i$. Meanwhile it suggests that in the sampling period the appearance order of those flow nodes including $i$ and its neighbors is cross, in other words, there are the inter-dependencies among these flows. The empirical network clustering coefficient results in Table\ref{tab:wide-result} denote that the ICMP has the best clustering feature in the investigated TLCNs, i.e., $C = 0.71$ where $C \in [0, 1]$. The second best is the TELNET that $C$ is 0.42. As we all know, the ICMP, as one of the most important Internet control protocols, is used to not only reply messages of many Internet protocols in addition to the ICMP request, but also test the connectivity between the client and server in a software. Therefore, a group of ICMP flows may be transmitted periodically to scan the online devices. Those periodic flows prefer to form the connections to each other. Additionally, for the Internet attack, especially multiple attackers to one victim (called as m-s attack) or one attacker to multiple victims (called as s-m attack), the clustering coefficient $c(i)$ of attack flow node is usually higher than others. Indeed, it has been verified by one s-m probe attack, as shown in Fig. \ref{sfig:net-probe-s-m}, that the $c(i)$ of attack flow node with red color is higher than background flow nodes.

\begin{figure}[htbp]
\subfloat[TELNET\label{sfig:richclub-telnet}]{%
  \includegraphics[width=.33\columnwidth]{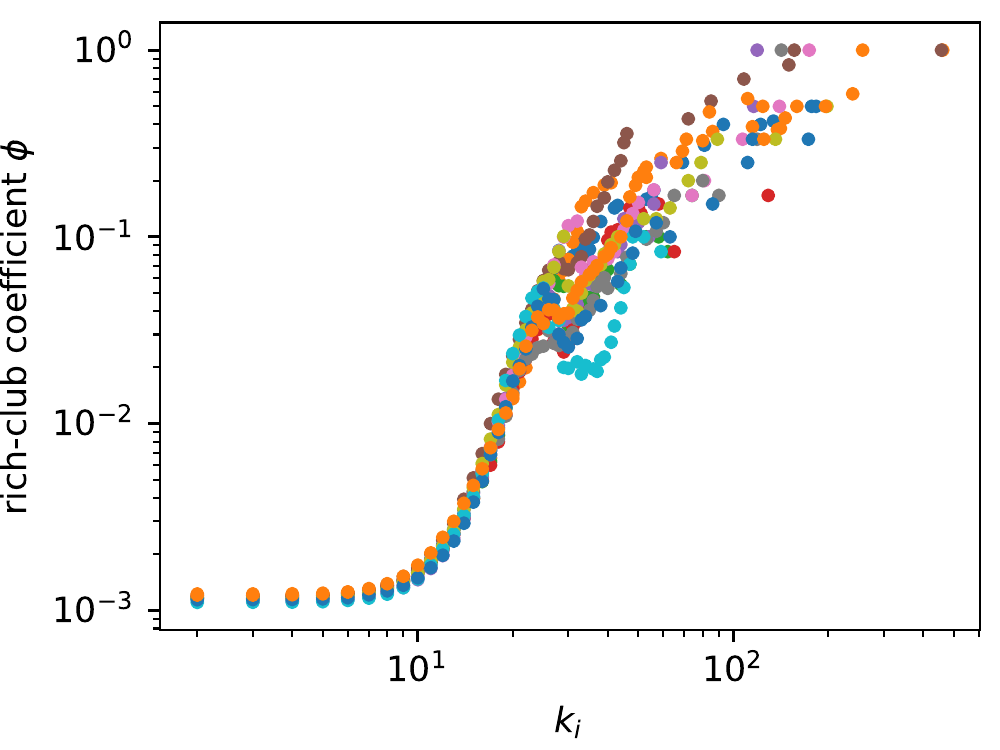}%
}\hfill
\subfloat[SMTP\label{sfig:richclub-smtp}]{%
  \includegraphics[width=.33\columnwidth]{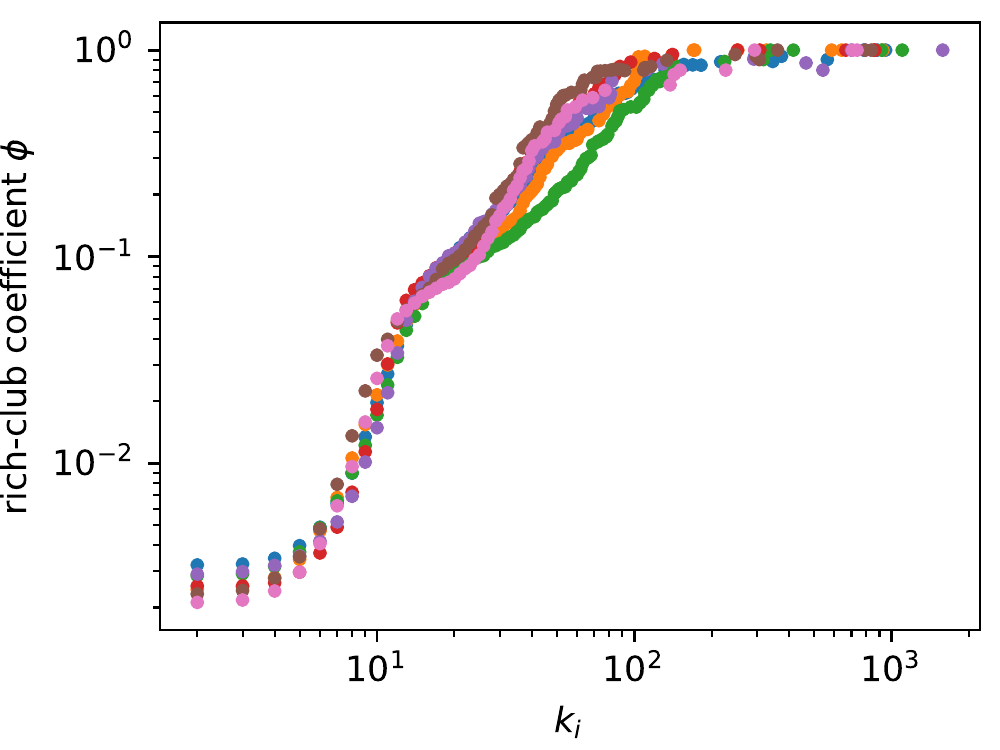}%
}\hfill
\subfloat[BT\label{sfig:richclub-bt}]{%
  \includegraphics[width=.33\columnwidth]{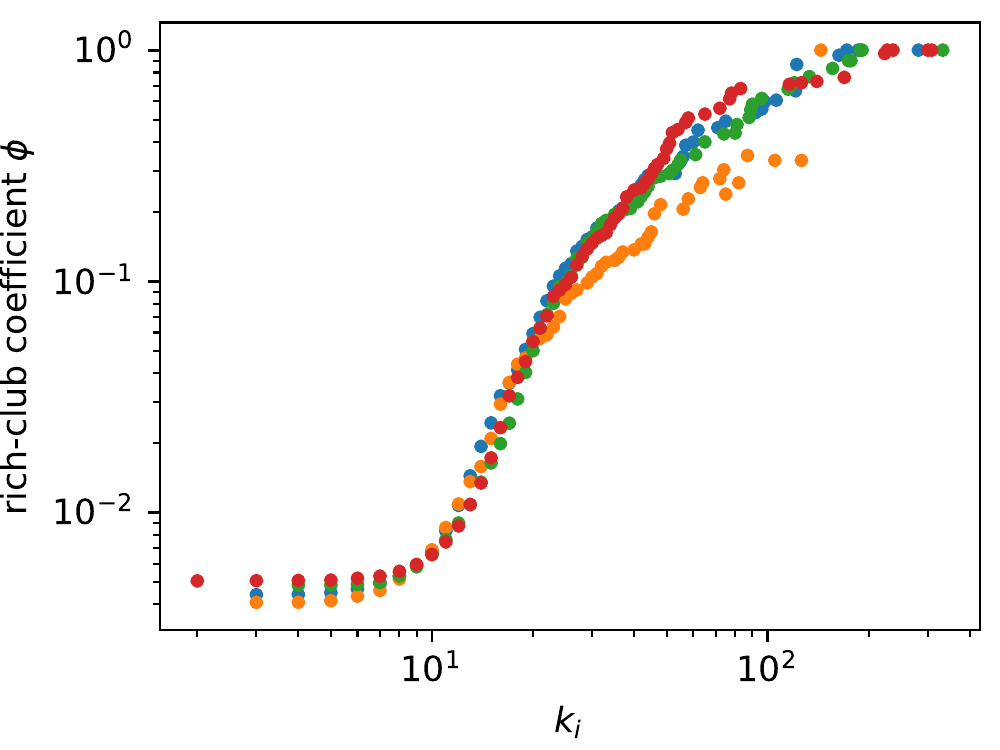}%
}\\
\subfloat[HTTP\label{sfig:richclub-http}]{%
  \includegraphics[width=.33\columnwidth]{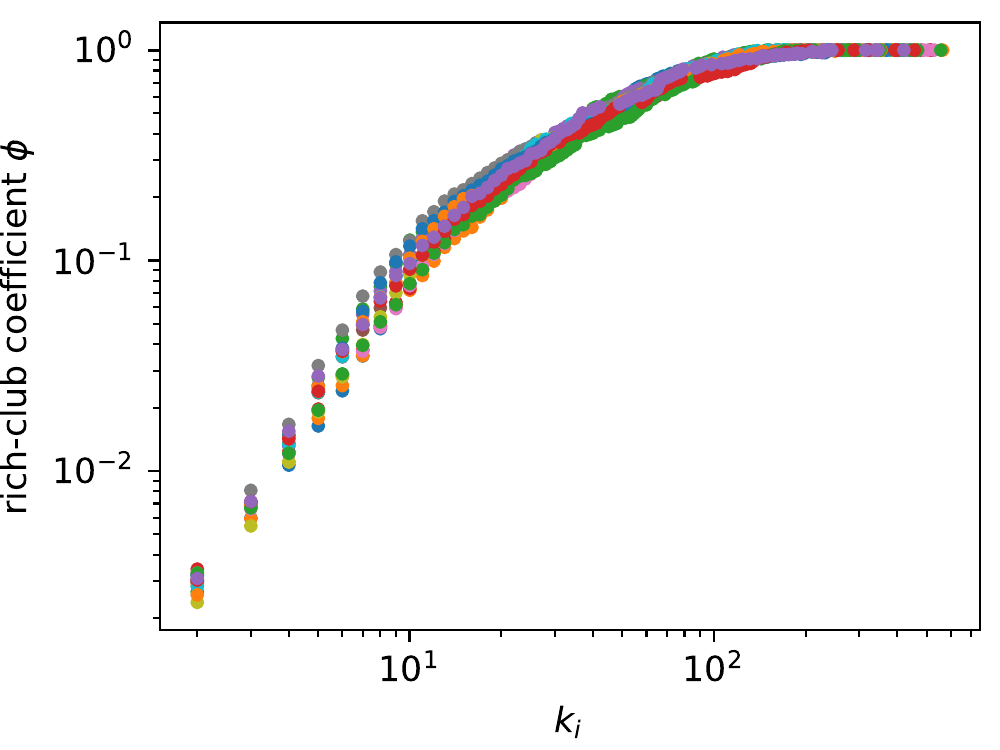}%
}\hfill
\subfloat[DNS\label{sfig:richclub-dns}]{%
  \includegraphics[width=.33\columnwidth]{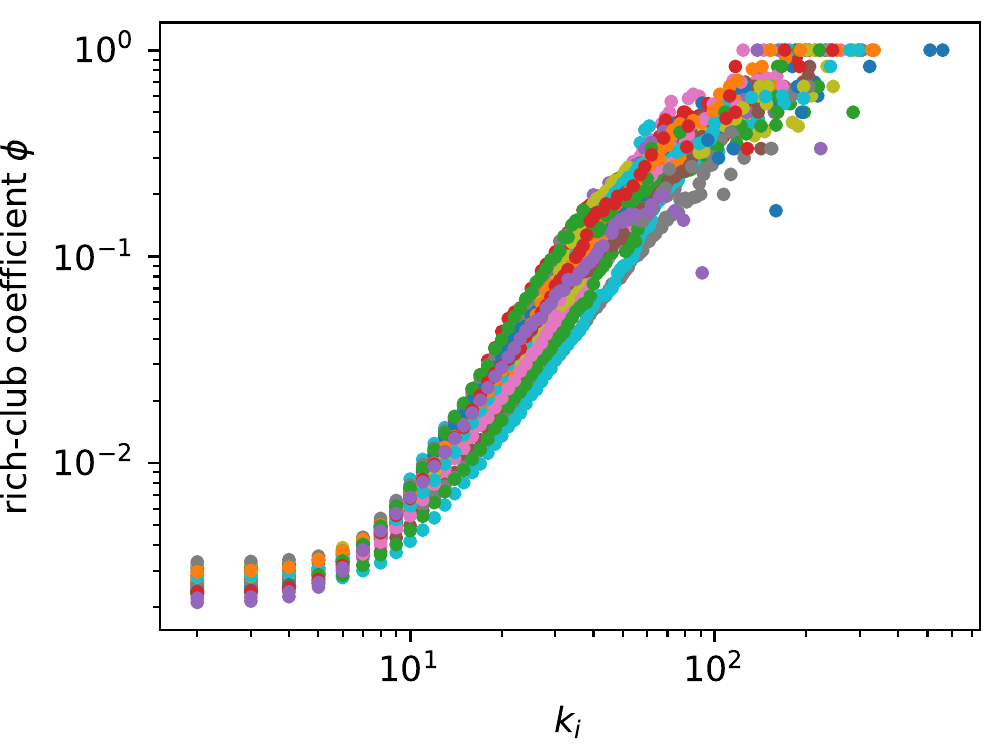}%
} \hfill
\subfloat[ICMP\label{sfig:richclub-icmp}]{%
  \includegraphics[width=.33\columnwidth]{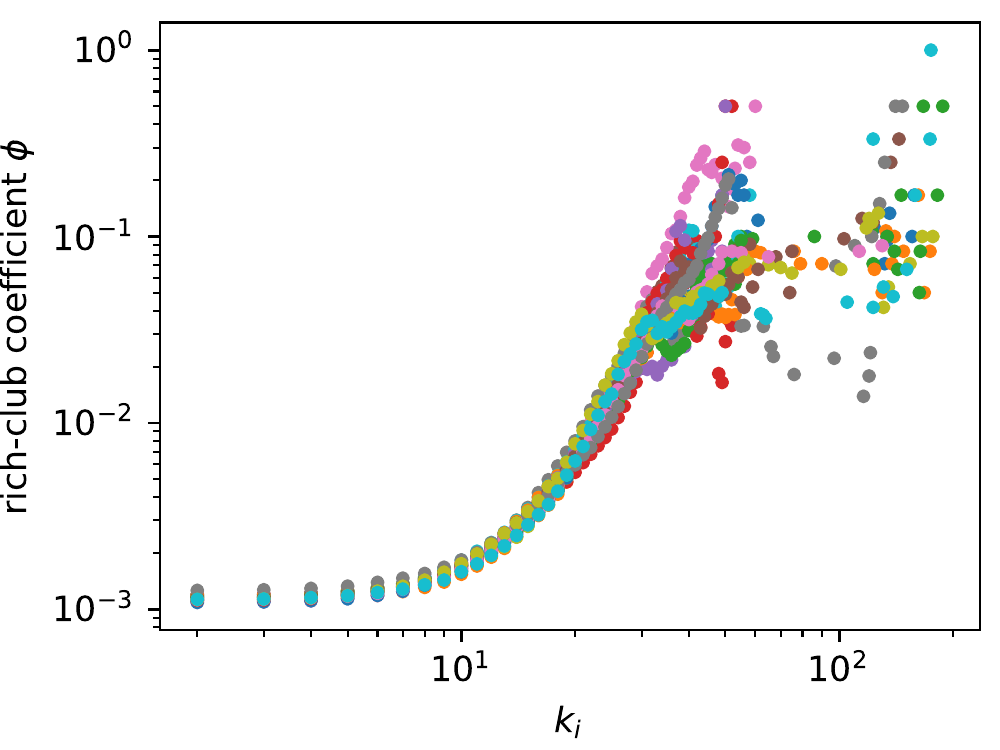}%
}
\caption{Distribution of rich-club coefficient $\phi$ of each TLCN.}
\label{fig:richclub-all}
\end{figure}

\textbf{Rich club:}\label{sec:rich-club}
Rich club phenomenon refers to the tendency of high degree nodes, the hubs of the network, to be very well connected to each other. Clearly, different from the clustering coefficient, the rich club mainly focuses on the transitivity of high degree nodes. In fact, those nodes with a large number of connections are much more likely to form tight and well-interconnected subnetwork than the low degree nodes. In this paper, the rich-club coefficient $\phi$ is introduced, $\phi(k) = {2E_{>k}}/{N_{>k}(N_{>k}-1)}$, to quantify the rich-club phenomenon, where $E_{>k}$ denotes the number of edges among the $N_{>k}$ nodes having higher degree than a given value $k$\cite{Colizza2006}. In brief, $\phi(k)$ measures the fraction of edges actually connecting those nodes out of the maximum number of edges they might possibly share. According to the Fig.\ref{fig:richclub-all}, it can be seen the SMTP (Fig.\ref{sfig:richclub-smtp}), BT (Fig.\ref{sfig:richclub-bt}) and HTTP (Fig.\ref{sfig:richclub-http}) have the remarkable rich-club phenomenon. The SMTP and BT not only have high MDR discussed above, but also have a well-interconnection in the nodes including MDR node and its neighbors. The high degree nodes in SMTP and BT represent the high-frequency communication flows between two busy hosts. From the perspective of application protocol, the busy hosts in SMTP are the servers of well-known mail service providers, and those in BT are the BT Tracker servers. Thus, among these hot-hosts there must be the continuous and high-frequency communications. Corresponding to the TLCN structure, the high degree nodes are well-interconnected to each other. For the HTTP, we believe that the high-frequency flows are mainly the caching flows between CDN server and Web server, and the response flows of static files such as the CSS and JavaScript. The CDN, as one of the most important components of the WWW, is enable to decrease the delay of HTTP request by distributed caching the contents of hot websites, such as the images and videos. Therefore, we usually can obtain the same response flows from CDN servers, while we send the HTTP requests with different destination IP address. Apparently, the caching flows become the high-frequency flows in the monitored network. Moreover, the developers would like to use some front-end libraries (e.g., JQuery, Bootstrap and React) in the web development. As a result, there are same flows from client to static libraries in spite of accessing different web systems.

Observing the distribution of the $\phi(k)$ in Fig.\ref{fig:richclub-all}, we found that the TELNET and ICMP have no the rich-club phenomenon. In order to answer why they are not fundamentally, the flow contents of the TELNET were analyzed. The results show that most of captured packets are SYN flood attack packets for multiple target subnets (i.e., 133.195.0.0/16, 133.58.0.0/16, 150.1.0.0/16, 152.231.0.0/16 and 202.0.0.0/8), and these packets only appear once or twice. For the ICMP traffic flows, most of them are mainly used to the host scan, which can be classified into two situations: intentional attack (e.g. ICMP flood attack) and the connectivity test between client and server. An important feature of the host scanning behavior is the subnet selection's randomness, but the local scan activities are regular. That is to say that the attacker will scan each of IP addresses incrementally after selects a subnet randomly. So the ICMP has a well interconnectivity in the local subnet flows but no high-frequency flows. For the reason, in ICMP there is a high clustering coefficient rather than the rich club phenomenon, as shown in Fig.\ref{sfig:net-icmp}.

\begin{figure}[htbp]
\subfloat[TELNET\label{sfig:pdf-telnet}]{%
  \includegraphics[width=.33\columnwidth]{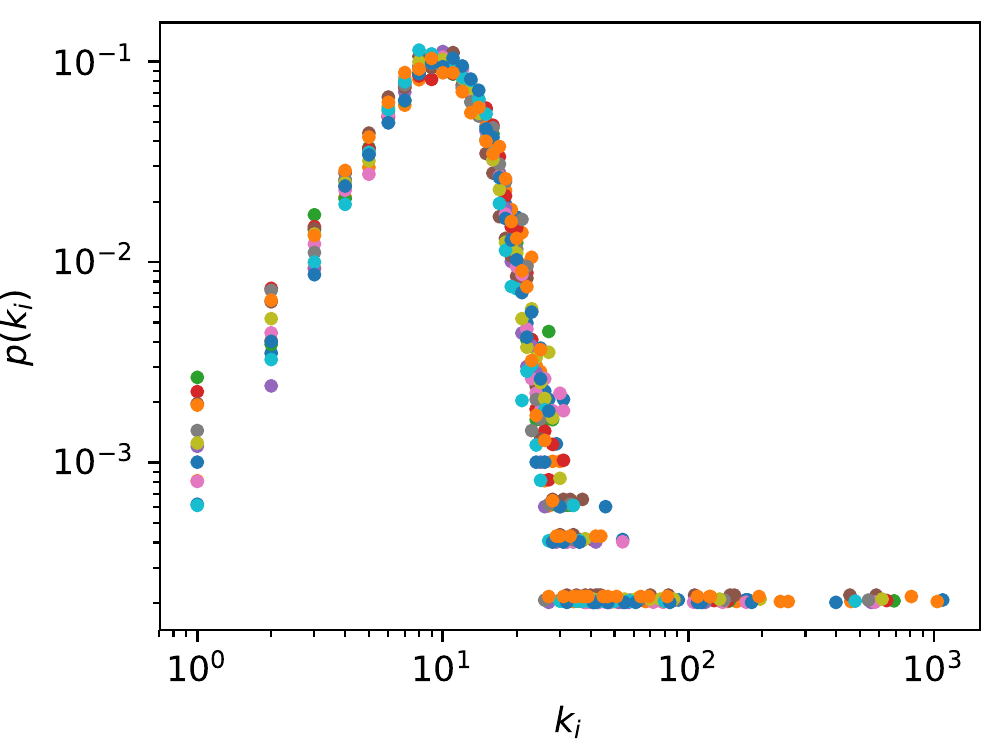}%
}\hfill
\subfloat[SMTP\label{sfig:pdf-smtp}]{%
  \includegraphics[width=.33\columnwidth]{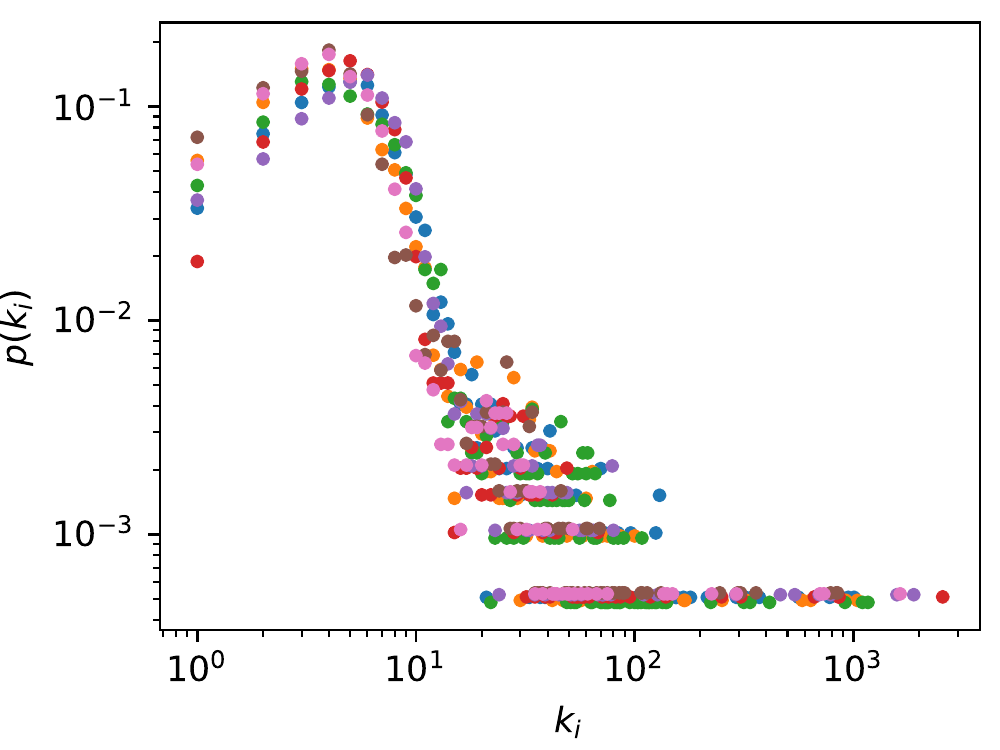}%
}\hfill
\subfloat[BT\label{sfig:pdf-bt}]{%
  \includegraphics[width=.33\columnwidth]{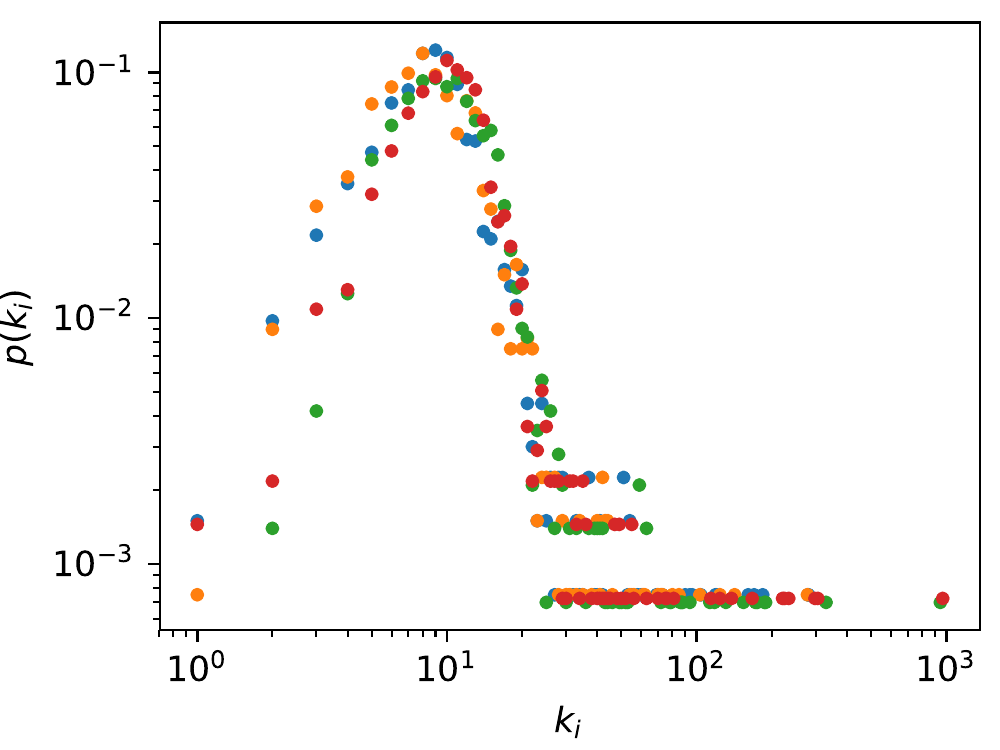}%
}\\
\subfloat[HTTP\label{sfig:pdf-http}]{%
  \includegraphics[width=.33\columnwidth]{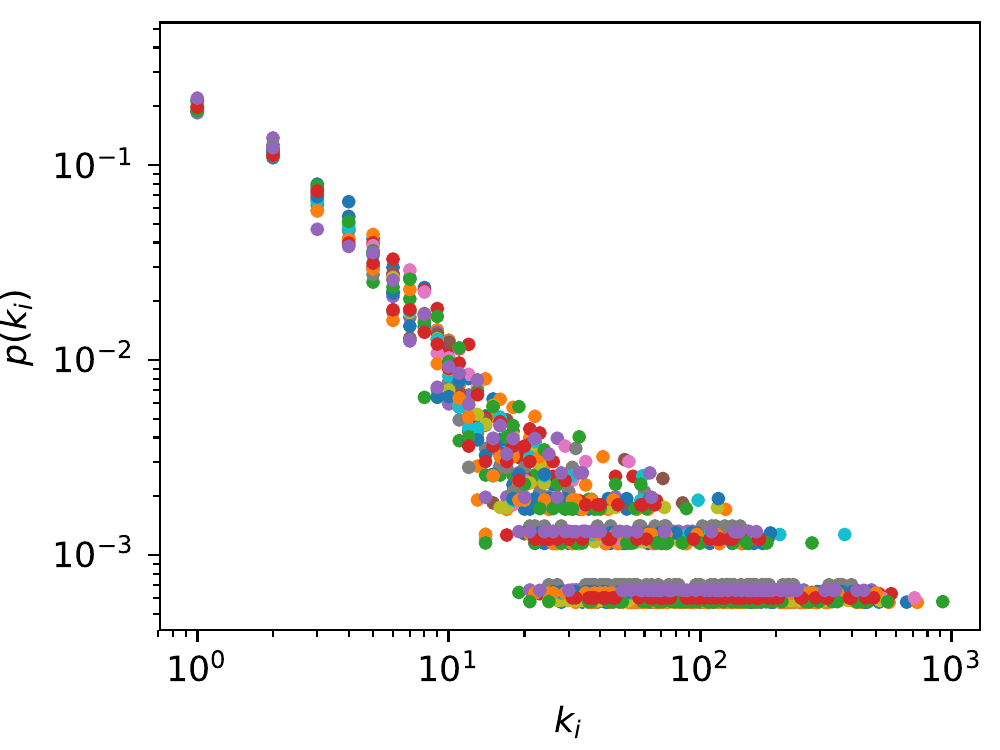}%
}\hfill
\subfloat[DNS\label{sfig:pdf-dns}]{%
  \includegraphics[width=.33\columnwidth]{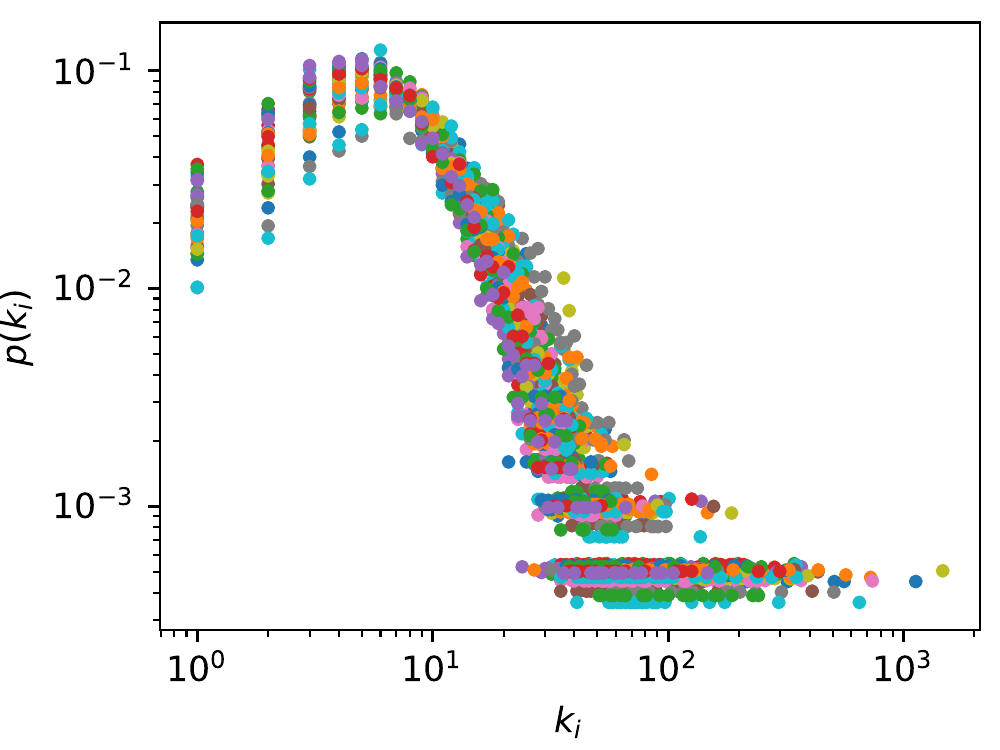}%
} \hfill
\subfloat[ICMP\label{sfig:pdf-icmp}]{%
  \includegraphics[width=.33\columnwidth]{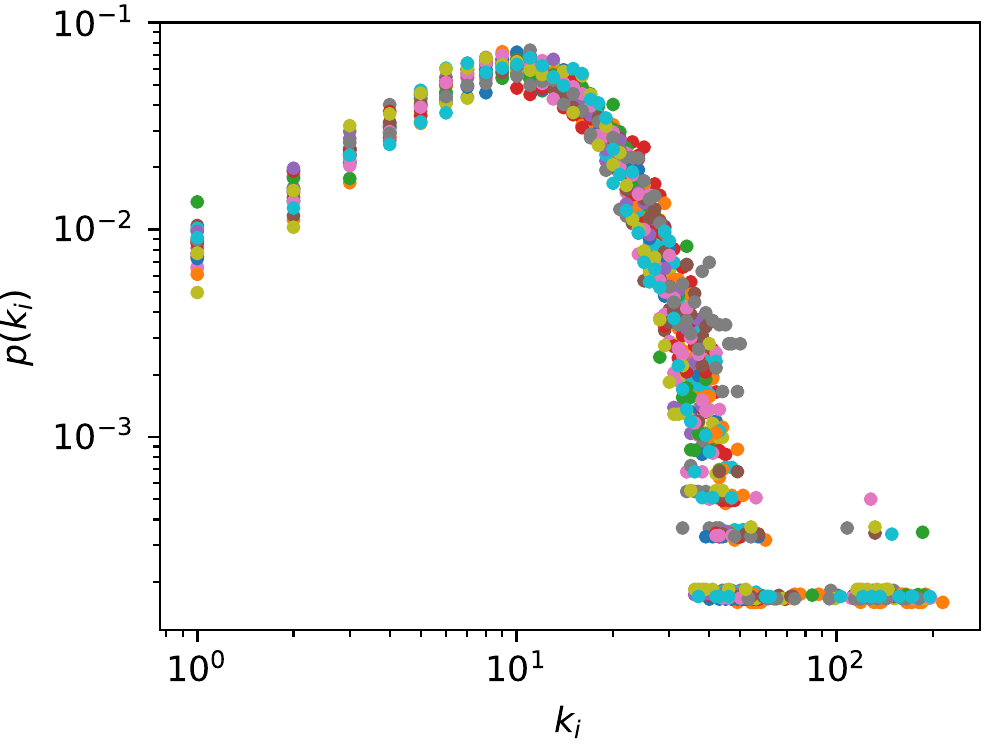}%
}
\caption{Degree Distribution $p(k_i)$ of each TLCN.}
\label{fig:pdf-all}
\end{figure}

\textbf{Degree distribution:}\label{sec:degree-disribution}
The degree distribution of any TLCN, i.e., the probability distribution of all node degrees over the whole network, can be represented by its corresponding marginal in-degree and out-degree distributions. In this work, we use the degree distribution of the undirected network in the same way as we discussed for the MDR. The empirical degree distributions of Fig.\ref{fig:pdf-all} indicate that (a) as with most Internet data\cite{Faloutsos1999,FAN2016327,CHEN2018191}, all TLCNs exhibit highly heavy-tailed distributions indicating the presence of the nodes with a much higher degree than most other nodes, and that (b) our measured quantities are stable over the temporal locality window, since the empirical distributions of all sampling TLCNs of each application are very similar.

Observing the node degree distributions, especially the low degrees, it can be seen that of TELNET, BT, and ICMP follow Poisson distributions. As discussed above, the clustering of the TELNET and ICMP is well so that the number of low degree nodes, less than 10 in this paper, increases with the increasing degree value. However, in the BT the slices of a share file are continuously transmitted in the P2P network so that the share flows are ease to form some clique structure (i.e., the full-connected subnetwork) in TLCN due to the interaction among the share flows. But the connections among those clique structures is few compared with the TELNET and ICMP. For this reason, the BT has low clustering coefficient.

Observing the Fig.\ref{fig:pdf-all}, the degree distributions of SMTP, HTTP and DNS can be well described by a power-law relationship of the form $P(k) \approx k^{-\lambda}$, with the power-law exponents $\lambda = 2.07$ for SMTP, $\lambda = 3.15$ for HTTP and $\lambda = 2.73$ for DNS and goodness-of-fit $R^2=0.87$, $R^2=0.97$ and $R^2=0.98$, respectively. Moreover, the degree distribution curves of SMTP and DNS exhibit a higher similarity, even though there is a slight difference between two exponent values. In fact, most of transmission flows in the SMTP and DNS mainly come from the hot hosts of well-known service providers. In other words, the transmission flows among these hot hosts dominate the SMTP or DNS applications.

\textbf{Assortativity coefficient:}\label{sec:assortativity-coefficient}
Assortativity is a preference for a network's nodes to attach to others that are similar in some way, and is expressed as a scalar value $\rho\in[-1, 1]$. According to the Table \ref{tab:wide-result}, it can be seen that the ICMP is assortative because high degree nodes are, on average, connected to other nodes with high degree and low degree nodes are, on average, connected to other nodes with low degree. The SMTP and HTTP are disassortative which, on average, high degree nodes are connected to nodes with low(er) degree and, on average, low degree nodes are connected to nodes with high(er) degree. Since the values of assortativity coefficient of TELNET, BT and DNS are very close to 0, we call them non-assortative networks. One one hand, the assortativity provides information about the structure of a network. For instance, when considering a random node $i$ of a assortative or disassortative network, we know what degree(s) to expect for the nodes connected to node $i$. When considering a random node $i$ of a non-assortative network, we have no expectation of the degree(s) of the nodes connected to node $i$ \cite{Solé2004}. On the other hand, it presents the dynamic and robustness of a network in flow attack (e.g., the traffic hijacking) and virus spread. In Ref.\cite{Wang2014}, the authors found the disassortative networks are vulnerable for flow attacks because the low degree nodes are easier to lose the connections to the network after the high degree nodes are attacked. But thanks to this structure feature, it is harder to spread the virus due to the low spreading rate of the disassortative network. As a result, the ICMP is more robust for flow attack but is easier to spread the ICMP virus, but the SMTP and HTTP are more vulnerable to flow attack but is hard to spread the SMTP and HTTP virus. In a way, Internet attacks improve the number of network traffic flows. The more flows mean more connections in the TLCN. So as we can see in the Fig.\ref{sfig:ctu-9-ac} the disassortative becomes more disassortative after it are attacked. Correspondingly, the the assortative network becomes more assortative.


\textbf{Path Length:}\label{sec:path-length}
The network path length exhibits the effectiveness of information transmission in a network. Recently, researchers have defined multiple metrics about the path length quantify problem \cite{Wang2017}. For instance, the shortest path length (SPL) is defined as the average number of steps along the shortest paths for all possible pairs of network nodes, and the network diameter is used to represent the maximum number of steps along the shortest paths. In general, the lower the value of SPL or network diameter is, the more efficient the network information transmission is. This is to say that there is a better network performance\cite{PAGANI2014248}. In the TLCN, the lower value of SPL or diameter implies that the interdependent relationships among traffic flows are stronger in the monitored network. However, the results about SPL and network diameter, especially the 2.59 and 8.0 for STMP, 2.67 and 5.6 for HTTP, and 3.79 and 10.77 for DNS, indicate that the TLCNs constructed by the strong-interaction flows have the better communication performance. Facing the Internet intentional attack the TLCN's SPL and network diameter will decrease because the attack increases the PPS of attacked routers. Indeed, it is verified in the CTU botnet attack, as shown in Fig.\ref{sfig:ctu-9-spl} and Fig.\ref{sfig:ctu-9-diameter}. Additionally, the network diameters of TELNET and ICMP, the 32.08 for TELNET and 23.45 for ICMP, are greater than others. As found above, the ruled flows in TLENET and ICMP are transmitted periodically so that the flow nodes within a time period are easy to form the well-interconnected subnetwork. But the connections among the subnetworks are so few that those weak connections increase the information transmission cost of the pairs of nodes that both are in different subnetworks.

\begin{table}[htbp]
\begin{center}
\caption{Measured characteristics(averages) for TLCNs generated within the applications in the WIDE dataset. Values in parenthesis provide the standard deviation for the measured quantity after generating each TLCNs.}\label{tab:wide-result}
\resizebox{\textwidth}{!}{
\begin{tabular}{l  c  c  c  c  c c c }
\hline
\hline
network & node & edge & MDR & Clustering & Assortativity & SPL & Diameter \\
\hline
TELNET & 4846(125) & 32191(1300) & 0.14(0.04) & 0.42(0.1) & -0.004(0.008) & 7.68(1.62) & 32.08(6.26) \\
STMP & 1959(74) & 12087(1903) & 0.74(0.32) & 0.04(0.02) & -0.2(0.03) & 2.59(0.2) & 8.0(1.53) \\
BT & 1370(47) & 10014(1357) & 0.45(0.27) & 0.24(0.09) & -0.04(0.04) & 3.27(0.46) & 10.25(1.71) \\
HTTP & 1611(98) & 9361(1198) & 0.34(0.07)  & 0.2(0.02) & -0.21(0.03) & 2.67(0.03) & 5.6(0.63) \\
DNS & 2072(210) & 13407(3148) & 0.19(0.13)  & 0.26(0.06) & 0.004(0.03) & 3.79(0.31) & 10.77(1.37)\\
ICMP & 5844(202) & 45122(3322) & 0.02(0.01)  & 0.71(0.04) & 0.32(0.23) & 7.15(1.1) & 23.45(4.08) \\
\hline
\hline
\end{tabular}
}
\end{center}
\end{table}

\section{Anomaly behavior analysis of TLCNs}
\label{sec:dynamic-tlcns}

\begin{figure}[htb]
\subfloat[DoS:processtable(s-s)\label{sfig:net-dos-s-s}]{%
  \includegraphics[width=.33\columnwidth]{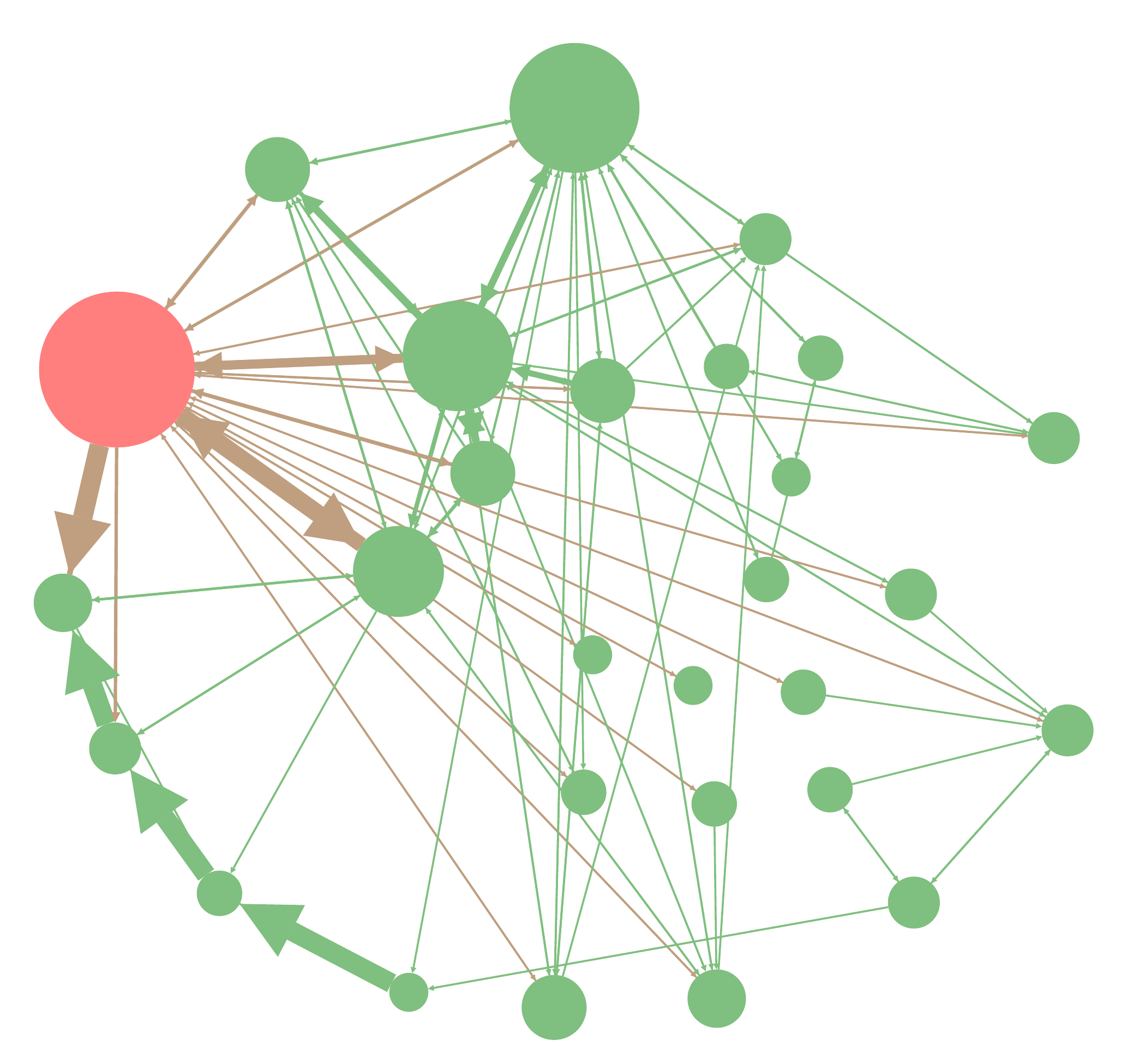}%
}\hfill
\subfloat[Probe:ipsweep(s-m)\label{sfig:net-probe-s-m}]{%
  \includegraphics[width=.33\columnwidth]{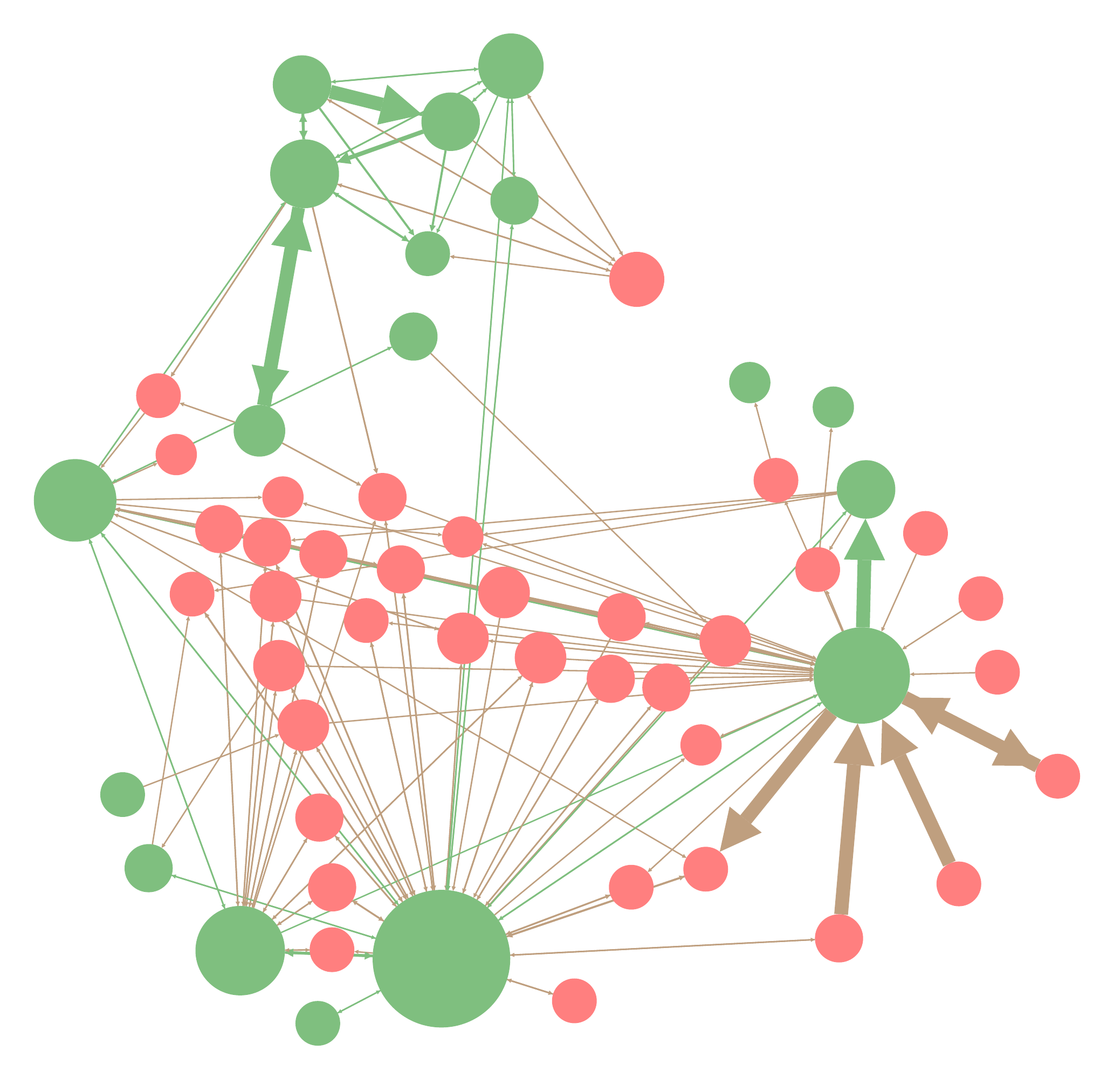}%
}\hfill
\subfloat[Probe:ipsweep(m-m)\label{sfig:net-probe-m-m}]{%
  \includegraphics[width=.33\columnwidth]{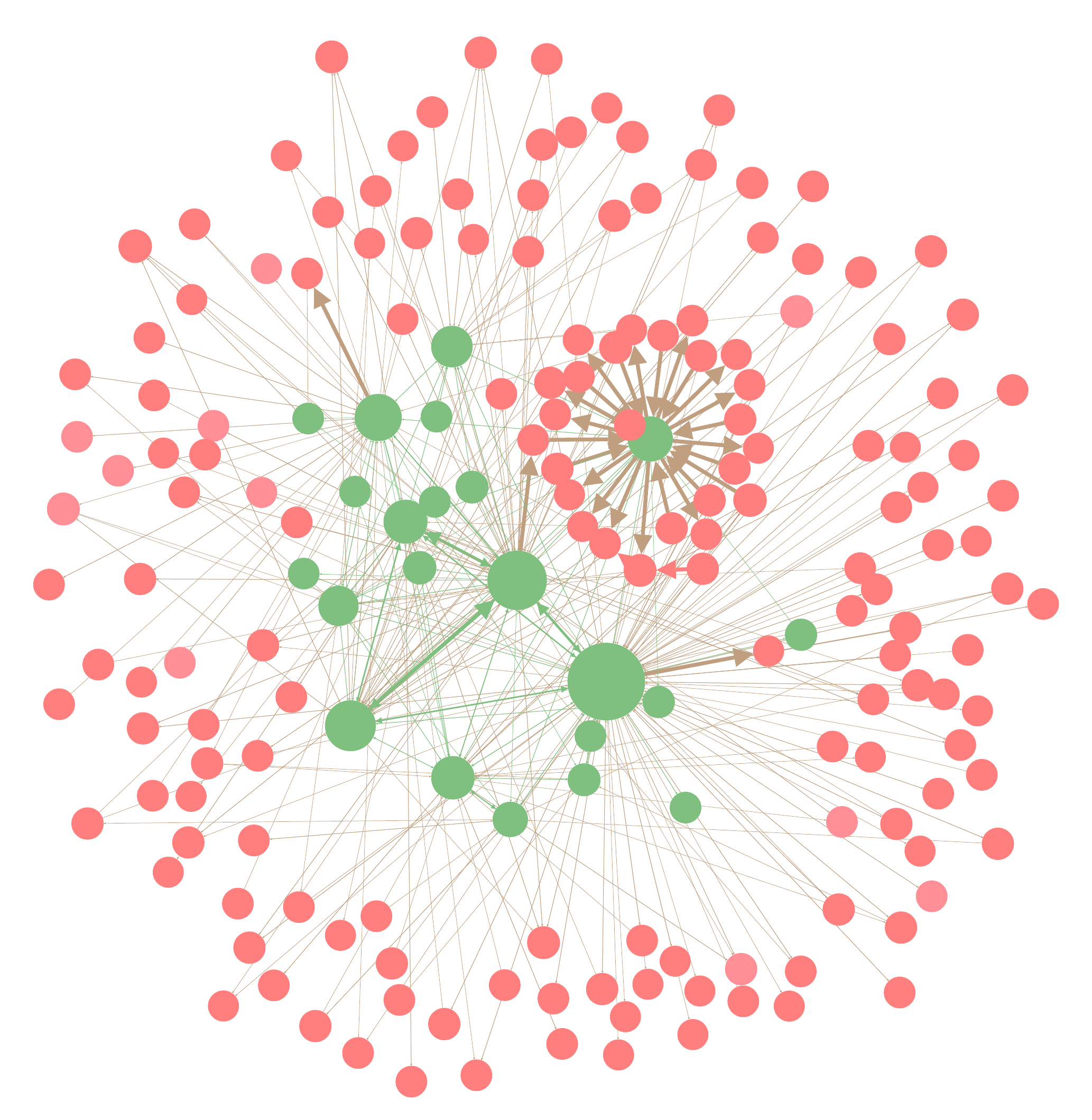}%
}
\caption{The visualization of TLCNs with the anomaly events from Darpa99 dataset. The green nodes and red nodes represent the normal flow and abnormal flow separately, and the size of the node is decided by its out-degree. Moreover, the width of edge denotes the edge weight. (a) the processtable attack from single attacker to single victim at 1999-03-30 17:49:15, (b) the ipsweep attack from single attacker to multiple victims at 1999-04-02 19:25:14 and (c) the ipsweep attack from multiple attacker to multiple victims at 1999-04-06 21:15:54.}
\label{fig:darpa-all}
\end{figure}

\begin{figure}[htb]
\captionsetup[subfigure]{labelformat=empty}
\subfloat{%
  \includegraphics[width=.49\columnwidth]{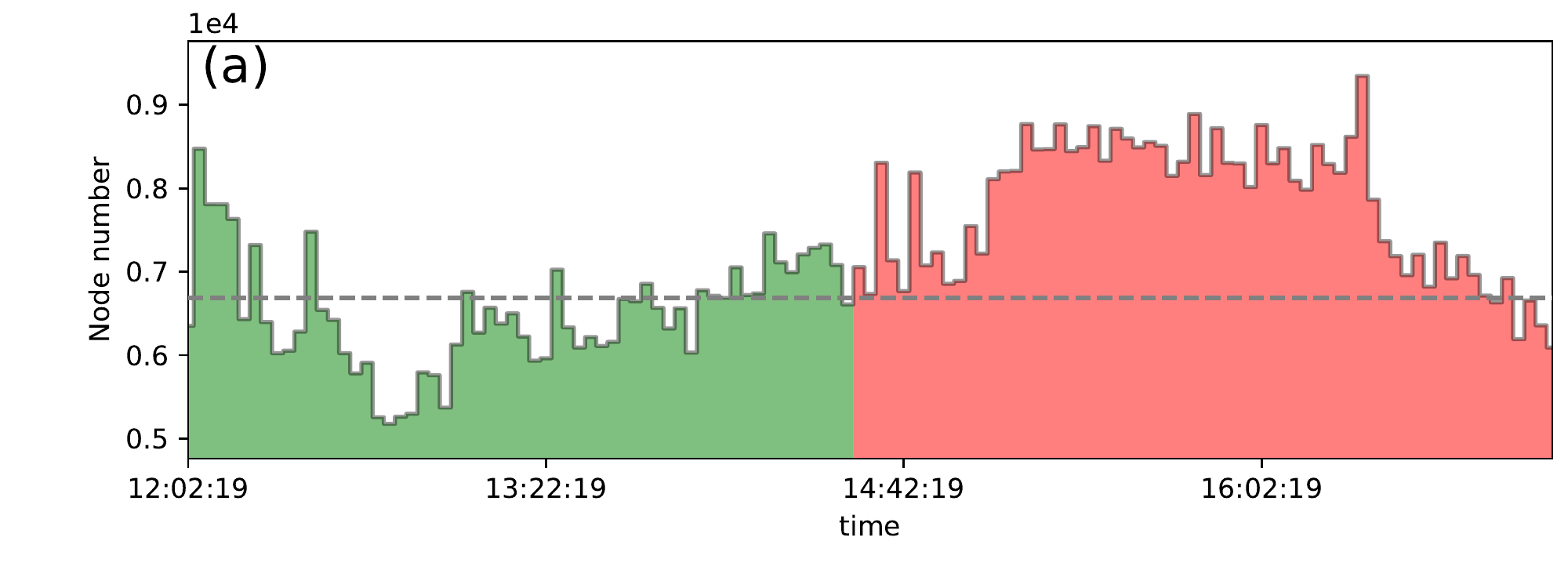}%
  \label{sfig:ctu-9-node}
}\hfill
\subfloat{%
  \includegraphics[width=.49\columnwidth]{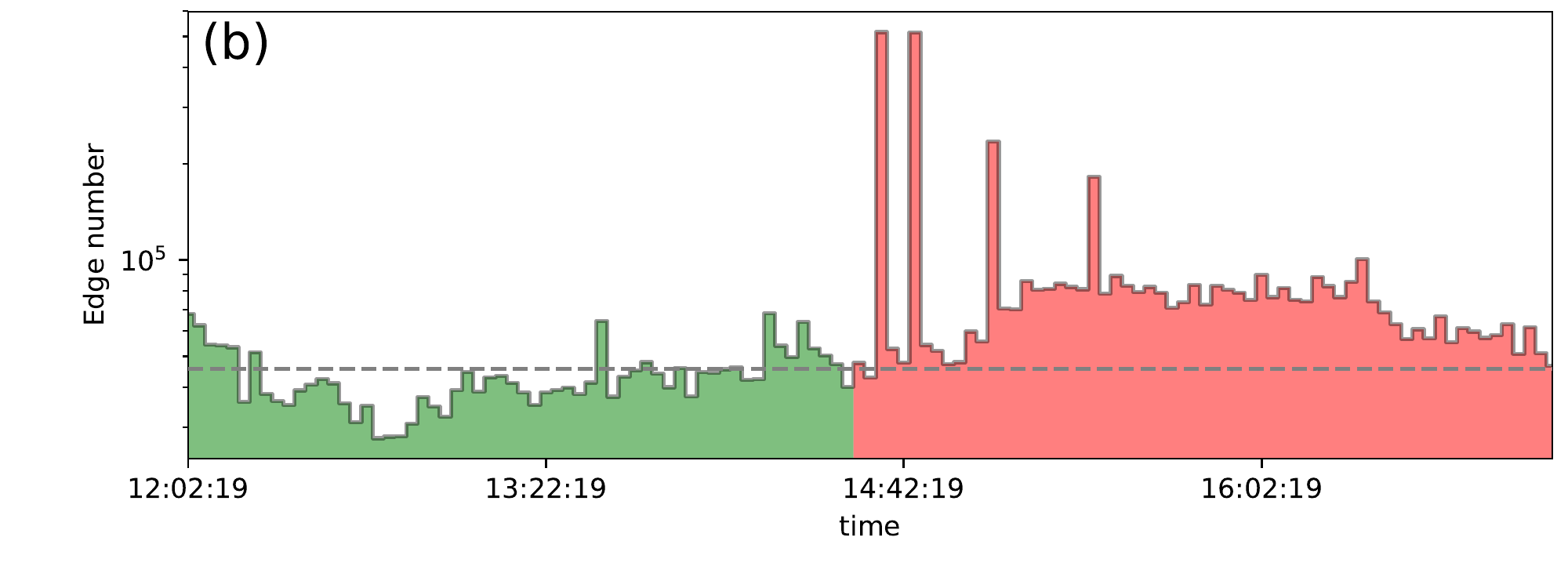}%
  \label{sfig:ctu-9-edge}
} \\
\subfloat{%
  \includegraphics[width=.49\columnwidth]{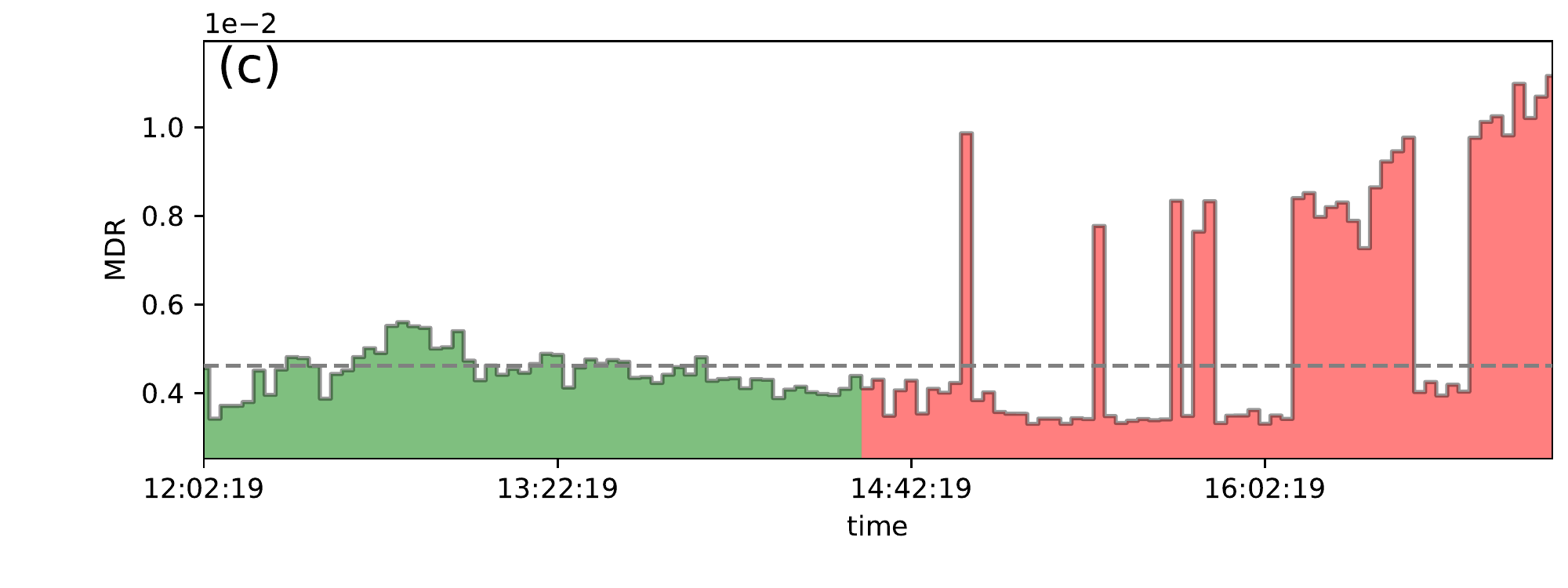}%
  \label{sfig:ctu-9-mdr}
}\hfill
\subfloat{%
  \includegraphics[width=.49\columnwidth]{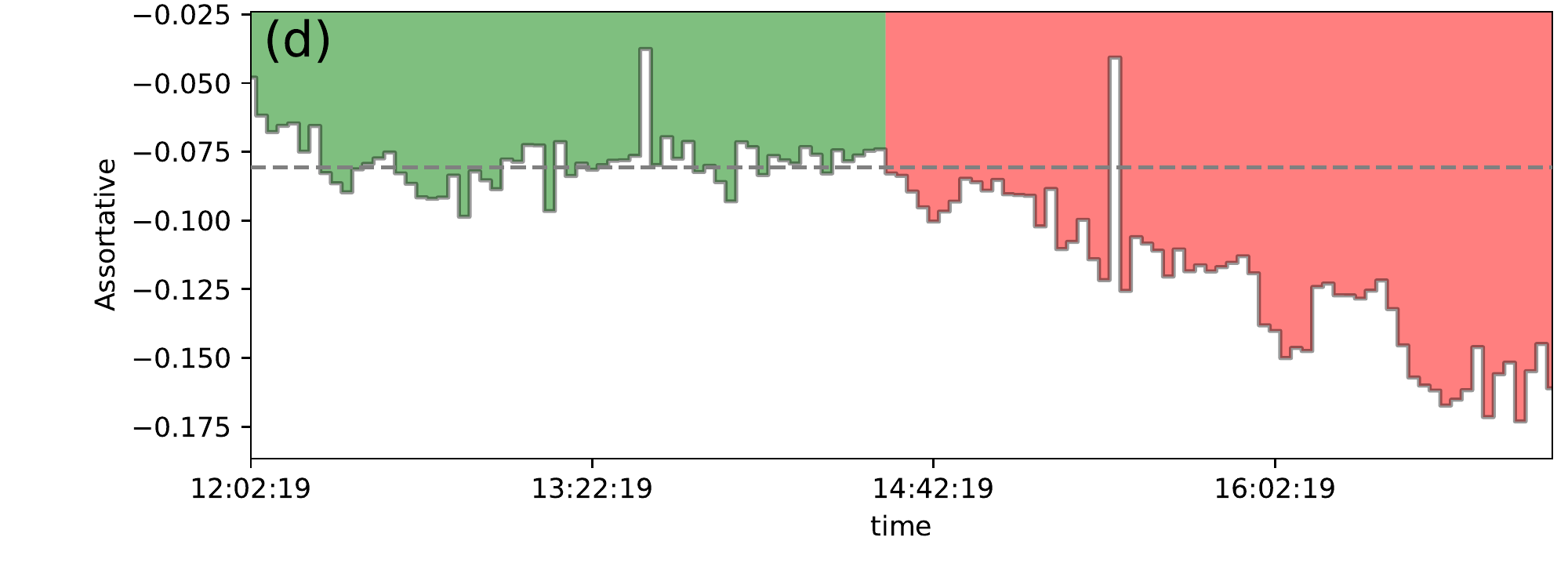}%
  \label{sfig:ctu-9-ac}
} \\
\subfloat{%
  \includegraphics[width=.49\columnwidth]{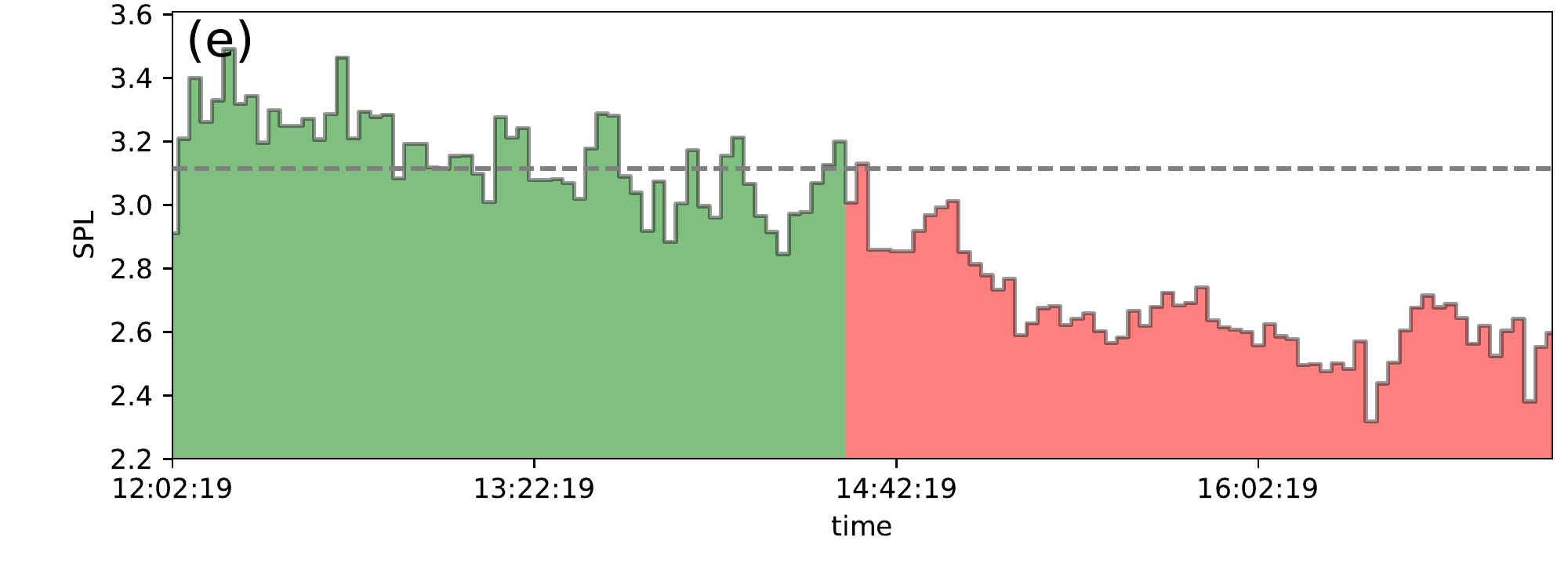}%
  \label{sfig:ctu-9-spl}
}\hfill
\subfloat{%
  \includegraphics[width=.49\columnwidth]{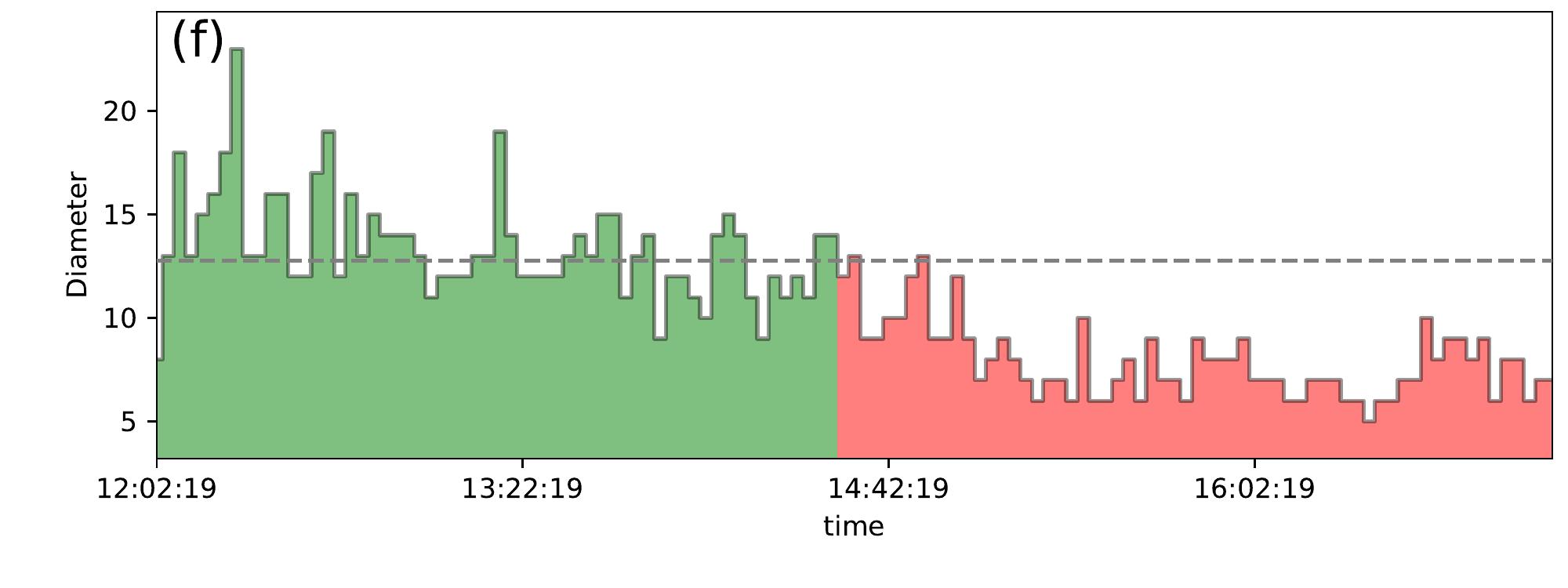}%
  \label{sfig:ctu-9-diameter}
}\hfill
\caption{Evolution of the characteristics of the DNS from the CTU dataset. (a) Node number vs. time, (b) Edge number vs. time, (c) MDR vs. time, (d) Assortativity coefficient vs. time, (e) SPL vs. time, (f) Diameter vs. time. For each sub-figure, the right green and left red areas denote the normal and abnormal traffic separately. The dashed lines represent the upper limit (i.e., (a), (b), (c)) or lower limit (i.e., (d), (e), (f)) of characteristic values of normal TLCNs with confidence interval $1-\alpha$ = 0.95}
\label{fig:ctu-9-all}
\end{figure}

In order to study the relationship between the dynamics of the TLCN  and network flow behavior, we investigated the network traffic flows with the anomaly events: the short-term attacks from DARPA dataset (Fig. \ref{fig:darpa-all}) and the long-term botnet from CTU dataset (Fi.g \ref{fig:ctu-9-all}). First, about the short-term attacks, three attack types are selected which include the DoS attack from single attacker to single victim (s-s DoS), TCP Probe attack from single attacker to multiple victims (s-m Probe) and ICMP Probe attack from multiple attackers to multiple victims (m-m Probe). The s-s DoS, as the Fig.\ref{sfig:net-dos-s-s} shown, causes sshd daemon on the victim to fork so many children that the victim can spawn no more processes. Clearly, the attack flow will become a maximum degree and betweenness node in the TLCN due to its high-frequency communication in the monitored network. But as this attack flow has no strong-interaction with others, its clustering coefficient is small, $c=0.17$. For the s-m Probe attack using the ipsweep tool (Fig.\ref{sfig:net-probe-s-m}), the attacker scans the subnet within a attack period by using the TCP flood. During the attack, after current attack flow is repeatedly sent 5 times, next flow will be done. Therefore, there are well local clustering for the attack flow nodes in TLCN, i.e. $c=0.45$ for the average of attack nodes and $c=0.29$ for the TLCN. But the degree and betweenness of those nodes is lower that non-attack nodes. In the Fig.\ref{sfig:net-probe-m-m}, the attackers used the ICMP probe attack based on the ipsweep tool. Different from the Fig.\ref{sfig:net-probe-s-m}, every attacker only sends once probe to each of hosts of its attacked subnet. In general, each attacker has its attack targets and attack frequency. But in this attack the attackers is only going on one attack period. As a result, 96.875\% of attack nodes is the unidirectional transit node that the values of their betweenness are 0 in directed weighted TLCN(Fig.\ref{sfig:net-probe-m-m}). Above all, the TLCN structure is a meaningful tool for describing different network attack patterns.

Then, the CTU bontnet traffic flows were adopted to observe the dynamical evolution of the TLCNs with the anomaly\cite{Wang2018}. The CTU dataset has a large capture of real botnet traffic mixed with normal traffic and background traffic by executing the Neris malware\footnote{\url{https://mcfp.felk.cvut.cz/publicDatasets/CTU-Malware-Capture-Botnet-50/}}. Based on TCP application protocol, the sampling TLCNs were constructed. Subsequently, 6 network characteristics were calculated and then plotted as a function of time, as shown in Fig.\ref{fig:ctu-9-all}. The network characteristic metrics include the node number, edge number, MDR, assortativity coefficient, SPL and network diameter. Following the method proposed in Ref.\cite{GARCIA2014100}, the real network states (i.e., normal or abnormal) at each time tick are labeled. Corresponding to the Fig.\ref{fig:ctu-9-all}, the left green area and right red area of each sub-figure represent the normal traffic and attack traffic. Clearly, the values of each of network characteristics from the attack traffic flows are very different from that of normal traffic flows. It suggests that the TLCN characteristic measurement will be a effective method for Internet anomaly detection.

According to the hoeffding's inequality\cite{Hoeffding1963}, we proposed the detection rules that the sampling TLCN is anomaly if the network characteristic value $c_i$ is lower than the $\psi$, or vice versa, in which the $\psi$ indicates the abnormal threshold of the system. So the value of the threshold $\psi$ is related to the performance of the anomaly detection. Given a confidence interval $\theta$, the threshold $\psi$ can be computed by $\psi = \mu + \lambda \sigma$, where $\mu$ and $\sigma$ denote the mean and the standard deviation of a characteristic time sires $C={c_1, c_2,\dots,c_n}$. And the notation $\lambda$ is the quantile of the normal distribution corresponding to the given confidence interval $\theta$\cite{meeker2014statistical}. In this paper, the $\varepsilon$ is 0.05, that is to say that confidence interval of a TLCN characteristic sequence is $1-\varepsilon = 0.95 $. Accordingly, the detection accuracy of TLCNs are 78.05\%, 85.37\%, 50.41\%, 81.30\%, 78.86\% and 78.86\%, respectively.

\section{Discussion}

One of the main goals of our work was to develop a network traffic flow profiling model to describe the complex interaction behaviors of large-scale network traffic flows. Based on the flow dependency or correlation in the temporal locality, we proposed the temporal locality complex network--TLCN. Then, we studied the statistical characteristics of static TLCNs based on Internet applications and analyzed the dynamics of the TLCNs with anomaly events.

On the one hand, according to observing the degree distribution in Fig.\ref{fig:pdf-all}, it can be seen that degree distributions of the TELNET, BT, and ICMP follow the Poisson distribution, however, that of the SMTP, HTTP and DNS are well described by a power-law relationship with the goodness-of-fit $R^2=0.87$, $R^2=0.97$ and $R^2=0.98$, respectively. According to the definition of scale-free network\cite{BARABASI1999}, it can be said that the SMTP, HTTP and DNS are scale-free networks. From the perspective of application protocol, we believe that SMTP, HTTP and DNS belong to the strong interaction type in traffic flows. Indeed, the dependencies among traffic flows are more common or even necessary for the Internet. But the important flows may be triggered frequently, such as the attachment transmission flows between the file storage servers in SMTP, the caching flows and static file flows in HTTP, and the query flows between the key DNS servers, by the important, unimportant or even occasional flows. To correspond to the TLCN construction, the above fact can be described by the preference attachment model that the high degree nodes have a greater probability of being selected to connect a new node. For the Poisson distribution's TLCN (e.g. the TLENET, BT and ICMP), the small-coefficient\cite{Humphries2008}, $\sigma = \frac{C/C_r}{SPL/SPL_r}$, was introduced to adjust whether it is random network or small-world network by comparing the clustering coefficient $C$ and shortest path length $SPL$ of a given network to an equivalent random network with same degree on average. According to computing the small-coefficients of the TLENET, BT and ICMP, i.e. 392.33, 48.87 and 647.34 which are significantly higher than expected by random chance, it can be found that the TLENET, BT and ICMP are small-world networks that have the lower shortest path length and the higher clustering coefficients. In light of the protocol definitions, the TELNET, BT and ICMP protocols mainly focus on the communication on a per flow. That is to say that the interactions among the flows are very weak. In fact, the developers would like to use them in some special scenes, such as the TELNET- and ICMP-based intentional attack, the BT spiders, and the large-scale Internet detection based on ICMP. In those cases, the executors usually send the request packets periodically to a target IP list. So the behavior of the traffic flows will become ruled in the monitored network. Hence, most of TLCNs based on TLENET, BT and ICMP applications are the regular small-world network. In summary, the TLCN has a good structure representing ability for different Internet applications.

On the other hand, the dynamic analysis of TLCNs not only shows the consistency between TLCN structure and flow behavior, but also provides the good ideas for finding the attack patterns and detecting the anomalous traffic. As far as we know, the intentional attack and malicious scanning are two of the most common of the Internet anomaly behaviors. A remarkable feature of both is traffic outbreak after the anomaly happens. Hence, the TLCN structure will have sudden changes in aspects of network characteristic metrics. But for the local structure consisting of the attack flow node and its neighbors, they are different with different attack patterns. For instance, discussed above, the maximum degree for the attack flow node in the s-s attack (\ref{sfig:net-dos-s-s}), the low degree and high clustering for attack flow nodes in the s-m attack (\ref{sfig:net-probe-s-m}), and the low degree and 0 betweenness for the m-m attack (\ref{sfig:net-probe-m-m}). If a deep machine model is developed to construct the relationship between the changes of local structure of TLCN and attack patterns, it would be a novel method in the study of attack pattern recognition. Frankly, the good detection for the anomalous traffic will provide the better base for attack pattern recognition. That is because it is more efficient for attack recognition if we could find the anomalous point firstly. Thus, the dynamic evolution results of TLCNs with the anomaly events, as shown in Fig.\ref{fig:ctu-9-all}, will inspire our future works that are based on the multi-characteristics of TLCN.

\section{Conclusions}

Based on the temporal locality of network traffic flow, that current flow triggers the further flows, the temporal locality complex network--TLCN is put forward to study the interaction behaviors among network traffic flows.

In the definition of the TLCN, 5 node filtering strategies and 2 edge filtering strategies proposed in this paper enhance the network presentation ability for flow interactions and is to monitor flow behaviors in different levels of network packets. However, in this paper we mainly focus on the flow behaviors in Internet applications by extracting application protocol- and application port-based weighted directed TLCNs. In TLCN, the only TLCN parameter--temporal locality window $\Delta w$--was studied against complex network structure and network throughput. The results show that there is a linear correlation between the $\Delta w$ and one-order complex network metric, and between the $\Delta w$ and network throughput, but the power correlation between the $\Delta w$ and two-order metric.

Then, we analyzed the static characteristics and the dynamic behaviors of the TLCN. On one hand, the complex network metrics including the node centrality (i.e., MDR), node connectivity (i.e., clustering coefficient and rich club), connection preference (i.e., degree distribution and assortativity coefficient) and network performance (i.e., SPL and network diameter) were introduced to study the structure feature of TLCNs with different Internet applications such as the application port-based TELENT, SMTP, BT HTTP, DNS and application protocol-based ICMP. The analysis results about statistical characteristic suggest that the TLCN constructed by the weak interaction flows is small-world network, such as the TELENT, BT and ICMP, and the TLCN from the strong interaction flows is scale-free network, such as the SMTP, HTTP, and DNS. Additionally, the high frequency flow prefer becoming the MDR node in TLCN, and the interdependency flows including strong interaction flows and weak interaction ruled flows have better local clustering. In brief, the TLCN structure can well present the interaction behaviors of large-scale network traffic flows. On the other hand, the anomaly behavior analysis of the TLCNs indicates that (1) the attacked TLCN has a remarkable structure feature for three attack patterns, i.e., the maximum degree for the attack flow node in the s-s attack, the low degree and high clustering for attack flow nodes in the s-m attack and the low degree and 0 betweenness for the m-m attack, and (2) the dynamic evolution of the TLCNs provides a novel idea for traffic anomaly detection. In this paper, one simple method proposed based on the hoeffding inequality is used to verify the effectiveness of anomaly detection using the statistical characteristic of complex network, i.e., the detection accuracy 78.05\% for node number, 85.37\% for edge number, 50.41\% for MDR, 81.30\% for assortative coefficient, 78.86\% for shortest path length and 78.86\% for network diameter.

In summary, complex network offers a new start to understand the statistical characteristic and dynamics in TLCN. Further research on TLCN will provide us more helpful conclusions or tools to explore the interaction behaviors in large-scale network traffic flows. All source codes of our method and the datasets used in this work are shared openly at \url{http://file.mervin.me/project/internet-tlcn}.

\section*{Acknowledgment}

This work was supported by the Fundamental Research Funds for the Central Universities (02190022117021, N171903002).

\textbf{Delarations of interest:} none.

\section*{References}

\bibliography{refs}

\end{document}